\input harvmac
\input amssym
\input epsf.tex


\figno=0

\def\alphadot{{\dot \alpha}}

\def\CM{{\cal M}}
\def\CN{{\cal N}}
\def\CB{{\cal B}}
\def\CC{{\cal C}}
\def\IL{\relax{\rm I\kern-.18em L}}
\def\IH{\relax{\rm I\kern-.18em H}}
\def\IR{\relax{\rm I\kern-.18em R}}
\def\IS{\relax{\rm \kern-.18em S}}
\def\IC{\, \relax{\rm I\kern-.56em C}}
\def\mathbbR{{\IR}}
\def\mathbbS{{\IS}}

\def\IZ{{\Bbb Z}}
\def\IC{{\Bbb C}}
\def\IS{{\Bbb S}}
\def\IR{{\Bbb R}}
\def\unit{\relax{\rm 1\kern-.26em I}}

\def\we{\wedge}
\def\te{\tilde{\epsilon}}

\def\fig#1#2#3{
\par\begingroup\parindent=0pt\leftskip=1cm\rightskip=1cm\parindent=0pt
\baselineskip=11pt \global\advance\figno by 1 \midinsert
\epsfxsize=#3 \centerline{\epsfbox{#2}} \vskip 12pt {\bf Fig.
\the\figno:} #1\par
\endinsert\endgroup\par
}
\def\figlabel#1{\xdef#1{\the\figno}}

\lref\KlebanovHB{
  I.~R.~Klebanov and M.~J.~Strassler,
  ``Supergravity and a confining gauge theory: Duality cascades and
  chiSB-resolution of naked singularities,''
  JHEP {\bf 0008}, 052 (2000)
  [arXiv:hep-th/0007191].
}

\lref\HKO{
  C.~P.~Herzog, I.~R.~Klebanov and P.~Ouyang,
  ``D-branes on the conifold and N = 1 gauge / gravity dualities,''
  arXiv:hep-th/0205100.
}

\lref\AharonyPP{
  O.~Aharony,
  ``A note on the holographic interpretation of string theory backgrounds  with
  varying flux,''
  JHEP {\bf 0103}, 012 (2001)
  [arXiv:hep-th/0101013].
}

\lref\GubserQJ{
  S.~S.~Gubser, C.~P.~Herzog and I.~R.~Klebanov,
  ``Symmetry breaking and axionic strings in the warped deformed conifold,''
  JHEP {\bf 0409}, 036 (2004)
  [arXiv:hep-th/0405282].
}

\lref\GubserTF{
  S.~S.~Gubser, C.~P.~Herzog and I.~R.~Klebanov,
  ``Variations on the warped deformed conifold,''
  Comptes Rendus Physique {\bf 5}, 1031 (2004)
  [arXiv:hep-th/0409186].
}

\lref\StrasslerQS{
  M.~J.~Strassler,
  ``The duality cascade,''
  arXiv:hep-th/0505153.
}

\lref\SeibergBZ{
  N.~Seiberg,
  ``Exact results on the space of vacua of four-dimensional SUSY gauge
  theories,''
  Phys.\ Rev.\ D {\bf 49}, 6857 (1994)
  [arXiv:hep-th/9402044].
}

\lref\SeibergPQ{
  N.~Seiberg,
  ``Electric - magnetic duality in supersymmetric nonAbelian gauge theories,''
  Nucl.\ Phys.\ B {\bf 435}, 129 (1995)
  [arXiv:hep-th/9411149].
}

\lref\IntriligatorAU{
  K.~A.~Intriligator and N.~Seiberg,
  ``Lectures on supersymmetric gauge theories and electric-magnetic  duality,''
  Nucl.\ Phys.\ Proc.\ Suppl.\  {\bf 45BC}, 1 (1996)
  [arXiv:hep-th/9509066].
}

\lref\Kachru{
  S.~Kachru, J.~Pearson and H.~L.~Verlinde,
  ``Brane/flux annihilation and the string dual of a non-supersymmetric  field
  theory,''
  JHEP {\bf 0206}, 021 (2002)
  [arXiv:hep-th/0112197].
}

\lref\Giddings{
  S.~B.~Giddings, S.~Kachru and J.~Polchinski,
``Hierarchies from fluxes in string compactifications,''
  Phys.\ Rev.\ D {\bf 66}, 106006 (2002)
  [arXiv:hep-th/0105097].
}

\lref\PT{
   G.~Papadopoulos and A.~A.~Tseytlin,
  ``Complex geometry of conifolds and 5-brane warped on 2-sphere,''
  Class.\ Quant.\ Grav.\  {\bf 18}, 1333 (2001)
  [arXiv:hep-th/0012034].
}

 \lref\KW{
  I.~R.~Klebanov and E.~Witten,
  ``Superconformal field theory on threebranes at a Calabi-Yau  singularity,''
  Nucl.\ Phys.\ B {\bf 536}, 199 (1998)
  [arXiv:hep-th/9807080].
}

\lref\HEK{
  C.~P.~Herzog, Q.~J.~Ejaz and I.~R.~Klebanov,
  ``Cascading RG flows from new Sasaki-Einstein manifolds,''
  JHEP {\bf 0502}, 009 (2005)
  [arXiv:hep-th/0412193].
}

\lref\BerensteinXA{
  D.~Berenstein, C.~P.~Herzog, P.~Ouyang and S.~Pinansky,
  ``Supersymmetry breaking from a Calabi-Yau singularity,''
  JHEP {\bf 0509}, 084 (2005)
  [arXiv:hep-th/0505029].
}

\lref\FrancoZU{
  S.~Franco, A.~Hanany, F.~Saad and A.~M.~Uranga,
  ``Fractional branes and dynamical supersymmetry breaking,''
  arXiv:hep-th/0505040.
}

\lref\BertoliniDI{
  M.~Bertolini, F.~Bigazzi and A.~L.~Cotrone,
  ``Supersymmetry breaking at the end of a cascade of Seiberg dualities,''
  Phys.\ Rev.\ D {\bf 72}, 061902 (2005)
  [arXiv:hep-th/0505055].
}

\lref\AffleckMK{
  I.~Affleck, M.~Dine and N.~Seiberg,
  ``Dynamical Supersymmetry Breaking In Supersymmetric QCD,''
  Nucl.\ Phys.\ B {\bf 241}, 493 (1984).
}
\lref\AffleckXZ{
  I.~Affleck, M.~Dine and N.~Seiberg,
  ``Dynamical Supersymmetry Breaking In Four-Dimensions And Its
  Phenomenological Implications,''
  Nucl.\ Phys.\ B {\bf 256}, 557 (1985).
}

\lref\Butti{
  A.~Butti, M.~Grana, R.~Minasian, M.~Petrini and A.~Zaffaroni,
``The baryonic branch of Klebanov-Strassler solution: A
supersymmetric family of SU(3) structure backgrounds,''
  JHEP {\bf 0503}, 069 (2005)
  [arXiv:hep-th/0412187].
}

\lref\KT{
  I.~R.~Klebanov and A.~A.~Tseytlin,
  ``Gravity duals of supersymmetric SU(N) x SU(N+M) gauge theories,''
  Nucl.\ Phys.\ B {\bf 578}, 123 (2000)
  [arXiv:hep-th/0002159].
}

\lref\KN{
  I.~R.~Klebanov and N.~A.~Nekrasov,
  ``Gravity duals of fractional branes and logarithmic RG flow,''
  Nucl.\ Phys.\ B {\bf 574}, 263 (2000)
  [arXiv:hep-th/9911096].
}

\lref\ArgyresEH{
  P.~C.~Argyres, M.~R.~Plesser and N.~Seiberg,
  ``The Moduli Space of N=2 SUSY {QCD} and Duality in N=1 SUSY {QCD},''
  Nucl.\ Phys.\ B {\bf 471}, 159 (1996)
  [arXiv:hep-th/9603042].
}

\lref\MN{
  J.~M.~Maldacena and C.~Nunez,
  ``Towards the large N limit of pure N = 1 super Yang Mills,''
  Phys.\ Rev.\ Lett.\  {\bf 86}, 588 (2001)
  [arXiv:hep-th/0008001].
}

\lref\Grana{
  M.~Grana, R.~Minasian, M.~Petrini and A.~Tomasiello,
  ``Generalized structures of N = 1 vacua,''
  arXiv:hep-th/0505212.
    M.~Grana, J.~Louis and D.~Waldram,
  ``Hitchin functionals in N = 2 supergravity,''
  arXiv:hep-th/0505264.
}

\lref\KWnew{
  I.~R.~Klebanov and E.~Witten,
  ``AdS/CFT correspondence and symmetry breaking,''
  Nucl.\ Phys.\ B {\bf 556}, 89 (1999)
  [arXiv:hep-th/9905104].
}

\lref\HKO{
  C.~P.~Herzog, I.~R.~Klebanov and P.~Ouyang,
  ``D-branes on the conifold and N = 1 gauge / gravity dualities,''
  arXiv:hep-th/0205100.
}

\lref\GKT{
  S.~S.~Gubser, I.~R.~Klebanov and A.~A.~Tseytlin,
  ``String theory and classical absorption by three-branes,''
  Nucl.\ Phys.\ B {\bf 499}, 217 (1997)
  [arXiv:hep-th/9703040].
}

\lref\Krasnitz{
  M.~Krasnitz,
  ``A two point function in a cascading N = 1 gauge theory from
  supergravity,''
  arXiv:hep-th/0011179.
}

\lref\GK{
  S.~S.~Gubser and I.~R.~Klebanov,
  ``Baryons and domain walls in an N = 1 superconformal gauge theory,''
  Phys.\ Rev.\ D {\bf 58}, 125025 (1998)
  [arXiv:hep-th/9808075].
}

\lref\KKLMMT{
  S.~Kachru, R.~Kallosh, A.~Linde, J.~Maldacena, L.~McAllister and S.~P.~Trivedi,
  ``Towards inflation in string theory,''
  JCAP {\bf 0310}, 013 (2003)
  [arXiv:hep-th/0308055].
}

\lref\KKLT{
  S.~Kachru, R.~Kallosh, A.~Linde and S.~P.~Trivedi,
  ``De Sitter vacua in string theory,''
  Phys.\ Rev.\ D {\bf 68}, 046005 (2003)
  [arXiv:hep-th/0301240].
}

\lref\Dvali{
  G.~R.~Dvali and S.~H.~H.~Tye,
  ``Brane inflation,''
  Phys.\ Lett.\ B {\bf 450}, 72 (1999)
  [arXiv:hep-ph/9812483].
}

\lref\Binetruy{
  P.~Binetruy and G.~R.~Dvali,
  ``D-term inflation,''
  Phys.\ Lett.\ B {\bf 388}, 241 (1996)
  [arXiv:hep-ph/9606342].
}

\lref\Halyo{
  E.~Halyo,
  ``Hybrid inflation from supergravity D-terms,''
  Phys.\ Lett.\ B {\bf 387}, 43 (1996)
  [arXiv:hep-ph/9606423].
}

\lref\BenvenutiWX{
  S.~Benvenuti, A.~Hanany and P.~Kazakopoulos,
  ``The toric phases of the Y(p,q) quivers,''
  JHEP {\bf 0507}, 021 (2005)
  [arXiv:hep-th/0412279].
}

\lref\WittenEP{
  E.~Witten,
  ``Branes and the dynamics of {QCD},''
  Nucl.\ Phys.\ B {\bf 507}, 658 (1997)
  [arXiv:hep-th/9706109].
}

\lref\DvaliXE{
  G.~R.~Dvali and M.~A.~Shifman,
  ``Domain walls in strongly coupled theories,''
  Phys.\ Lett.\ B {\bf 396}, 64 (1997)
  [Erratum-ibid.\ B {\bf 407}, 452 (1997)]
  [arXiv:hep-th/9612128].
}

\lref\Kovner{
  A.~Kovner, M.~A.~Shifman and A.~Smilga,
  ``Domain walls in supersymmetric Yang-Mills theories,''
  Phys.\ Rev.\ D {\bf 56}, 7978 (1997)
  [arXiv:hep-th/9706089].
}

\lref\Hanany{
  A.~Hanany and E.~Witten,
  ``Type IIB superstrings, BPS monopoles, and three-dimensional gauge
  dynamics,''
  Nucl.\ Phys.\ B {\bf 492}, 152 (1997)
  [arXiv:hep-th/9611230].
}

\lref\Acharya{
  B.~S.~Acharya and C.~Vafa,
  ``On domain walls of N = 1 supersymmetric Yang-Mills in four dimensions,''
  arXiv:hep-th/0103011.
}

\lref\BinetruyHH{
  P.~Binetruy, G.~Dvali, R.~Kallosh and A.~Van Proeyen,
  ``Fayet-Iliopoulos terms in supergravity and cosmology,''
  Class.\ Quant.\ Grav.\  {\bf 21}, 3137 (2004)
  [arXiv:hep-th/0402046].
}

\lref\IK{See, for example, I. R. Klebanov, talk at {\it Strings 2004}.}

\lref\AharonyZR{
  O.~Aharony, A.~Buchel and A.~Yarom,
  ``Holographic renormalization of cascading gauge theories,''
  Phys.\ Rev.\ D {\bf 72}, 066003 (2005)
  [arXiv:hep-th/0506002].
}

\lref\BurgessIC{
  C.~P.~Burgess, R.~Kallosh and F.~Quevedo,
  ``de Sitter string vacua from supersymmetric D-terms,''
  JHEP {\bf 0310}, 056 (2003)
  [arXiv:hep-th/0309187].
}

\lref\Dine{
  M.~Dine, N.~Seiberg and E.~Witten,
  ``Fayet-Iliopoulos Terms In String Theory,''
  Nucl.\ Phys.\ B {\bf 289}, 589 (1987).
}


\Title{\vbox{\baselineskip12pt \hbox{hep-th/0511254}
\hbox{PUPT-2183} \hbox{ITEP-TH-62/05} }}
{\vbox{ \centerline{
On the Moduli Space of the Cascading }
\smallskip
\centerline{ $SU(M+p)\times SU(p)$ Gauge Theory
 } }}
\smallskip
\centerline{Anatoly Dymarsky,$^{1}$
Igor R. Klebanov,$^{1}$ and Nathan Seiberg$^2$}
\bigskip

\centerline{\it $^1$ Joseph Henry Laboratories, Princeton
University} \centerline{\it Princeton, New Jersey 08544, USA}
\smallskip

\centerline{\it $^2$ Institute for Advanced Study} \centerline{\it
Princeton, New Jersey 08540, USA}

\bigskip
\noindent
 We carry out a thorough analysis of the moduli space of the
cascading gauge theory found on $p$ D3-branes and $M$ wrapped
D5-branes at the tip of the conifold. We find various mesonic
branches of the moduli space whose string duals involve the warped
deformed conifold with different numbers of mobile D3-branes. The
branes that are not mobile form a BPS bound state at threshold. In
the special case where $p$ is divisible by $M$ there also exists a
one-dimensional baryonic branch whose family of supergravity
duals, the resolved warped deformed conifolds, was constructed
recently. The warped deformed conifold is a special case of these
backgrounds where the resolution parameter vanishes and a $\IZ_2$
symmetry is restored. We study various brane probes on the
resolved warped deformed conifolds, and successfully match the
results with the gauge theory. In particular, we show that the
radial potential for a D3-brane on this space varies slowly,
suggesting a new model of D-brane inflation.

\bigskip
\Date{November 2005}
\vfil\eject

\newsec{Introduction}

Consideration of $p$ D3-branes at the tip of the conifold leads to
the duality conjecture \KW\ relating type IIB string theory on
$AdS_5\times T^{11}$ to a superconformal $SU(p)\times SU(p)$ gauge
theory. Addition of $M$ D5-branes wrapped over the two-cycle of
$T^{11}$ deforms the gauge group to $SU(M+p)\times SU(p)$ \GK\ and
breaks the conformal invariance, producing a logarithmic running
of the gauge couplings \KN. This theory exhibits a ``duality
cascade'' \refs{\KT,\KlebanovHB} where along the RG flow $p$
repeatedly drops by $M$ units as a result of the duality of
\SeibergPQ.

The complete and non-singular supergravity dual of the cascading
gauge theory, the warped deformed conifold, was found in
\KlebanovHB. In the infrared it exhibits confinement and chiral
symmetry breaking, while in the UV there is a logarithmic running
of coupling constants and a duality cascade. In the absence of
extra branes, this background is dual to the $SU(M(k+1))\times
SU(Mk)$ theory at a special $\IZ_2$-symmetric point on the
baryonic branch ${\cal A} {\cal B}=const$ where the two baryonic
condensates are equal, $|{\cal A}|= |{\cal B}|$
\refs{\KlebanovHB,\AharonyPP,\GubserQJ}. The cascading gauge
theory has a pseudoscalar Goldstone mode of the spontaneously
broken $U(1)_{baryon}$, and its massless scalar superpartner
\AharonyPP. The supergravity duals of these modes were found in
\GubserQJ. The scalar zero-mode, which produces a small motion
along the baryonic branch was found with the help of the
Papadopoulos-Tseytlin ansatz \PT\ generalizing the $SO(4)$ and
$\IZ_2$ symmetric warped deformed conifold to include a breaking
of the $\IZ_2$. A general analysis of the supersymmetry conditions
for this ansatz led to a derivation \Butti\ of coupled first-order
equations describing the entire baryonic branch of confining
vacua. This family of {\it resolved warped deformed conifolds} is
then readily constructed through numerical integration of the
equations of \Butti\ subject to the requirement
 that at large radius they asymptote to the cascading solution of
\KT.

On the $SU(N_1)\times SU(N_2)$ gauge theory side, an analysis of
various branches of the moduli space was begun in \KlebanovHB, and
continued in \StrasslerQS. In this paper we carry out a complete
analysis and compare it successfully with the dual string theory.
The branches are labelled by two integers. One of them,
$r=1,...,M$, is associated with a spontaneous breaking of the
$\IZ_{2M}$ R-symmetry to $\IZ_2$.  The other, $l=0,...,k=[p/M]$,
has the following interpretation:
 $M(l+1)$ D5-branes and $lM$
anti-D5-branes which wrap the two-cycle of $T^{1,1}$ form a bound
state at threshold.  The remaining $p-lM$ D3-branes are free to
move on the deformed conifold whose deformation parameter
$\epsilon$ depends on $r$ and $l$.  In the special case where $p$
is a multiple of $M$ there exists a branch with no mobile
D3-branes, $l=k$.  This branch breaks the baryon number symmetry
of the problem and will be referred to as a baryonic branch.  It
is reminiscent of the baryonic branch of \ArgyresEH.  This picture
of the moduli space follows from a careful field theory analysis
using standard techniques (for a review, see e.g.\
\IntriligatorAU), but a few interesting subtleties which have so
far been ignored turn out to be quite important.

For the specific example of the baryonic branch, which exists only
for $N_1= M(k+1),\ N_2=Mk$, we carefully impose the boundary
conditions on the numerical solution, and show that they lead to a
constant tension of the BPS domain wall of \refs{\DvaliXE\Kovner-\WittenEP}
along the entire branch.
We also calculate the tensions of various other objects, the
confining string, the D3 and anti-D3 branes, and find that they
blow up far along the branch. We find that small departures from
the $\IZ_2$ symmetric point create a small potential for a
D3-brane which depends on the radial coordinate of the classical
solution. This suggests a string theoretic mechanism for brane
inflation where a D3-brane rolls towards smaller radius on a
resolved warped deformed conifold embedded into a flux
compactification. This approach is similar to that of KKLMMT
\KKLMMT, but instead of an anti-D3 brane uses a Fayet-Iliopoulos
parameter $\xi$ \GubserTF\ to resolve the warped
deformed conifold and generate a potential for the
D3-brane. Our proposal is therefore similar to the D-term
inflation mechanism of \refs{\Binetruy,\Halyo}. D-terms
also play an important role in string theoretic constructions
involving D7-branes and D3-branes \BurgessIC.

The paper is organized as follows. In section 2 we review the
gauge theory  and its symmetries. In section 3 we state our main
result for the quantum structure of the moduli space and discuss
its D-brane interpretation. Section 4 is devoted to the classical
analysis of the mesonic and the baryonic branches of the moduli
space. In section 5 we analyze the $SU(p+M)\times SU(p)$ gauge
theory with $p<M$, and find the quantum deformation of the
classical mesonic branches. In section 6 we study $p=M$ at the
quantum level, and find a mesonic and a baryonic branch. In
sections 7 and 8 we study the quantum moduli spaces for $p=M+1$
and $p> M+1$, respectively. In section 9 we discuss how the
different theories and different branches are related by Higgsing
and duality transformations. This leads to non-trivial consistency
checks of our quantitative results. In section 10 we summarize our
results on the gauging of $U(1)_{baryon}$ and turning on the
Fayet-Iliopoluos parameter $\xi$.

In section 11 we compare various gauge theory and corresponding
string theory objects. We match the BPS and non-BPS domain walls
present in the gauge theory with D5-branes and NS5-branes wrapped
over the three-sphere at the bottom of the deformed conifold. We
present a general argument showing that the tension of BPS domain
walls is independent of the moduli.  We also comment on how the
tensions of confining strings, glueballs, and solitonic strings
depend on the continuous parameter $g_s M$ present in the
cascading gauge theory. In sections 12 and 13 we review and
present some new results on the resolved warped deformed
conifolds, which are supergravity duals of the baryonic branch. We
solve the equations derived in \Butti, while carefully imposing
the boundary conditions at large radius. In section 14 we check
the consistency of our numerical solutions by showing that the BPS
domain wall tension is constant along the branch; we also study
the tensions of confining strings and of anti-D3-branes along the
baryonic branch. In section 15 we study the potential generated
for D3-branes and suggest a string theoretic implementation of the
D-term inflation. Some possible extensions of our work are
mentioned in the Discussion. Appendices A and B contain some
further details about the resolved warped deformed conifold.

\newsec{The gauge theory}

In this section
we consider the gauge dynamics of the supersymmetric field theory
with gauge group
 \eqn\gaugeg{SU(N_1=M+p)\times SU(N_2=p)}
with $p\ge 0$ (clearly, $N_1\ge N_2$).  We parameterize it as
 \eqn\para{N_1=(k+1)M + \tilde p \qquad ; \qquad  N_1=k M + \tilde
 p \qquad ; \qquad \tilde p=0,...,M-1}
We add matter fields
 \eqn\matter{\eqalign{
 &A_{\alpha i}^a ~~ {\rm in}~~ ({\bf N_1},\overline{\bf N_2})\ ,\cr
 & B_{\alphadot a} ^i~~ {\rm in}~~ (\overline{\bf N_1},{\bf N_2})}}
($\alpha,\alphadot=1,2$, $i=1,...,N_1$, $a=1,...,N_2$), and a tree
level superpotential
 \eqn\superp{W_0=h \Tr_a \det_{\alpha\dot\alpha} A_\alpha
 B_\alphadot= h\left(A_{1 i}^a B_{1 b} ^i
 A_{2 j}^b B_{2 a} ^i-A_{1 i}^a B_{2 b} ^i
 A_{2 j}^b B_{1 a} ^i\right)\ .}
This gauge theory
describes $N_1$ D5-branes and $N_2$ anti-D5-branes
wrapping the collapsed $\Bbb S^2$ at the singularity of the
conifold $\CC_0$. $\CC_0$ is parameterized by four complex numbers
$z_{\alpha\alphadot}$ subject to the equation $\det_{\alpha
\alphadot} z_{\alpha\alphadot}=0$.

The $SU(N_1)$
($SU(N_2)$) gauge theory has $2N_2$ ($2N_1$) flavors.  Therefore,
if the superpotential \superp\ is ignored,
the instanton factors of these two gauge groups are
 \eqn\instfac{\Lambda_1^{3N_1-2N_2}=\Lambda_1^{3M+p} \qquad ;
 \qquad \Lambda_2^{3N_2-2N_1}=\Lambda_2^{p-2M}}

Let us discuss the global symmetries of this theory.  Clearly,
there is an $SU(2)\times SU(2)$ symmetry which acts on the indices
$\alpha$ and $\alphadot$.  The global Abelian symmetry can be
analyzed in the basis
 \eqn\basissy{
 \matrix{
 & U(1)_A & U(1)_B & U(1)_R\cr
 A&1&0&1\cr
 B&0&1&1\cr
 h&-2&-2&-2\cr
 \Lambda_1^{3M+p}&2p &2p &2(M+p) \cr
 \Lambda_2^{p-2M}&2(M+p) &2(M+p) &2p}}
The exact symmetry of the system is the subgroup of \basissy\
which is not broken by nonzero $h$ and the anomalies.  It is
$U(1)_{baryon} \times \IZ_{2M}$. $U(1)_{baryon} $ is generated by
the difference of the $U(1)_A$ and $U(1)_B$ generators, and
$\IZ_{2M}$ is an R-symmetry which is generated by $A\to e^{2 \pi i
\over 2M} A$, $B\to e^{2 \pi i \over 2M} B$, and $\theta \to e^{2
\pi i \over 2M} \theta$.  Below we will also discuss the effect of
gauging $U(1)_{baryon}$ and adding a Fayet-Iliopoulos parameter
$\xi$.  On the string theory side this happens when the gauge
theory is embedded into a flux compactification with a compact
Calabi-Yau space.

We will find it convenient to form the following combinations
of the parameters
 \eqn\conop{ \matrix{
 & U(1)_A & U(1)_B & U(1)_R\cr
 L_1(M,p)=h^p\Lambda_1^{3M+p}&0 &0 &2M \cr
 L_2(M,p)=h^{M+p}\Lambda_2^{p-2M}& 0&0 &-2M\cr
 I(M,p)=L_1(M,p)L_2(M,p) &0&0&0}}
which do not transform under $U(1)_A \times U(1)_B$.

The combination $I(M,p)$ in \conop\ which is invariant under all
the symmetries is dimensionless, and therefore it is invariant
under the renormalization group.  It has a natural interpretation
in the brane system as the instanton factor of the type IIB string
theory
 \eqn\invcom{I(M,p)=L_1(M,p)L_2(M,p)=h^{M+2p}
 \Lambda_1^{3M+p}\Lambda_2^{p-2M}=e^{2\pi i \tau}}
One aspect of this interpretation is that $I(M,p)= e^{2\pi i
\tau}$ is the amplitude of a type IIB D-instanton.
It is related to two fractional D-instantons on the conifold corresponding
the instantons of the two gauge groups, $SU(N_1)$ and $SU(N_2)$.
This explains why $I(M,p)$ includes
the instanton factors of the two gauge groups.

The ratio $L_1(M,p)/L_2(M,p)$ is determined by the NS-NS and RR
two form potentials through the two sphere: \eqn\twopot{
{L_1(M,p)\over L_2(M,p)}\sim \exp \left [ {1\over \pi
\alpha'}\int_{\Bbb S^2} (B_2/g_s + i C_2) \right ] \ . } In the
conformal case $M=0$, similar relations between gauge theory and
string theory parameters were proposed in \KW.

\newsec{Summary of the gauge theory results}

To facilitate the reading of the paper, we summarize here our
conclusions about the moduli space of vacua of the quantum field
theory. For more details, see sections 4 - 11.

Our conclusion will be that the classical and quantum moduli spaces
of vacua are quite different.  The quantum moduli space is
 \eqn\quanm{\oplus_{r=1}^M\oplus _{l=0}^k
 Sym_{p-lM}(\CC_{r,l})}
Here $\CC_{r,l}$ is the deformed conifold which is smooth. It is
described by an equation in four complex variables
$z_{\alpha\alphadot}$
 \eqn\defconz{\det_{\alpha\alphadot} z_{\alpha\alphadot}=
 \epsilon}
The $\CC_{r,l}$ arising on different branches of \quanm\
have different deformation parameters $\epsilon_{M,P}(r,l)$ (see
below).  In the special case of $\tilde p=0$, the last term
$Sym_{p-kM=\tilde p=0}(\CC_{r,l=k})$ is replaced by $\IC$.

The sum over $r$ in \quanm\ reflects the spontaneous breaking of
the global $\IZ_{2M}$ R-symmetry to $\IZ_2$, which is achieved through
gluino condensation in a low energy $SU(M)$ subgroup.  This
$SU(M)$ group arises differently on different branches in
\quanm\ and in different regions of the moduli space on the same
branch.  Sometimes it is a nontrivial subgroup of the microscopic
$SU(N_1)\times SU(N_2)$, while in other cases it involves dual gauge
groups.  Its instanton factor is
 \eqn\scalel{
 \Lambda(M,p,l)^{3M} \sim\, h^{p+l(M+2p)}
 \Lambda_1^{(3M+p)(l+1)}\Lambda_2^{(p-2M)l}=L_1(M,p) I(M,p)^l}
and its gluino condensation leads to the low energy superpotential
 \eqn\suplg{W(M,p,l,r)=M \left(\Lambda(M,p,l)^{3M}\right)^{1\over
 M} = M \left(L_1(M,p)^{3M}\right)^{1\over M}
 \left(I(M,p)\right)^{l\over M}}
Above we used the invariant combination $I(M,p)$ of \invcom, and
we suppressed the phase $e^{2\pi i r \over M}$ which arises from
the branch of the fractional power and leads to the $r$
dependence. Note that the $l$ dependence is only through the power
of $I$. Below we will see in detail how the superpotential \suplg\
is generated; it arises differently on different branches.

In our case the parameters $z_{\alpha\alphadot}$ arise from
eigenvalues of the matrix
 \eqn\cmdeff{\sqrt h \CM_{\alpha \alphadot b}^a = \sqrt h
 A_{\alpha i}^a B_{\alphadot b}^i}
where the factor of $\sqrt h$ is introduced to simplify the
equations. We will see that the deformation parameter $\epsilon$
depends on the branch in \quanm,
 \eqn\epsilonl{
 \epsilon_{M,p}(r,l)\sim\,  \left(\Lambda(M,p,l)^{3M}
 \right)^{1\over M} \sim\, \epsilon_{M,p}(r,l=0) I(M,p)^{l\over M}}
The label $r$ in $\epsilon_{M,p}(r,l)$ is the branch of the
fractional power $e^{2\pi r i\over M}$ (to simplify
the equation we suppress this factor).

In terms of the D-brane interpretation, the space
$Sym_{p-lM}(\CC_{r,l})$ describes $p-lM$ pairs of
D5-anti-D5-branes each forming a D3-brane which leaves the tip of
the conifold and is free to move in its bulk after it is deformed
to $\CC_{r,l}$. The remaining $(l+1)M$ D5-branes and $lM$
anti-D5-branes form {\it a bound state at threshold.}  The
deformation by \epsilonl\ with its $M$ branches labelled by $r$ in
\quanm\ is generated by the strong coupling $SU(M)$ dynamics with
scale $\Lambda(M,p,l)$ of the D5-branes near the tip.

The branch of the moduli space in \quanm\ with
$Sym_{p-lM}(\CC_{r,l})$ will turn out to have the massless photons
of $U(1)^{p-lM-1}$ -- one fewer than the number of mobile
D3-branes. If $U(1)_{baryon}$ is gauged, then there is one more $U(1)$
factor and the low energy spectrum is that of $p-lM$ multiplets of
$\CN=4$. This is consistent because the global $U(1)_{baryon}$
symmetry is not broken on all these branches.

An important exception to this is the last term $Sym_0(\CC_{r,l})$
which appears only when $\tilde p=0$. In this case $U(1)_{baryon}$
is broken and hence $Sym_0(\CC_{r,l})$ should be replaced by a
copy of $\IC$.  This is the baryonic branch.  If $U(1)_{baryon}$
is gauged, then it is Higgsed, and this branch is lifted
\refs{\GubserQJ,\GubserTF}. In this case $\IC$ is replaced by a
point, and the pattern is more uniform. In section 10 we will
discuss the effect of turning on a Fayet-Iliopoulos parameter
$\xi$.

\newsec{Classical flat directions}

In this section we examine the classical moduli space of vacua.
The D-term equations set
 \eqn\cdterm{\eqalign{
 &\sum_\alpha A_\alpha A_\alpha^\dagger - \sum_\alphadot
 B_\alphadot^\dagger B_\alphadot=
{{\cal U} \over p } \unit_p \cr
 &\sum_\alpha A_\alpha^\dagger A_\alpha  - \sum_\alphadot
 B_\alphadot B_\alphadot^\dagger =
{  {\cal U}  \over M+ p} \unit_{M+p}
 } }
where $\unit_p$ and $\unit_{M+p}$ are $p\times p$ and $(M+p)\times
(M+p)$ unit matrices, and
 \eqn\ztwoop{ {\cal U}= \Tr \left (
 \sum_\alpha A_\alpha A_\alpha^\dagger - \sum_\alphadot
 B_\alphadot^\dagger B_\alphadot \right )
 \ . }
In the quantum theory, ${\cal U}$ is an operator, whose
expectation value labels different ground states. Gauging
$U(1)_{baryon}$ sets ${\cal U} =0$ in \cdterm. However, in this
case a Fayet-Iliopoulos term $\xi$ can be added, and then \cdterm\
has
 \eqn\xiu{  {\cal U} = \xi\ .}

The solutions of these equations together with the F-term
equations depend on the values of $p$ and $M$.  We find two kinds
of classical solutions which we refer to as mesonic and baryonic.

\subsec{Mesonic flat direction}

We always have the mesonic flat directions (up to gauge
transformations)
 \eqn\claflaf{\eqalign{
 &A_\alpha=\pmatrix{
 A_{\alpha 1}^1 &&&&&&&&&&\cr
 &A_{\alpha 2}^2 &&&&&&&&&\cr
 && A_{\alpha 3}^3 &&&&&&&&\cr
 &&&. &&&&&&&\cr
 &&&&.&&&&&&\cr
 &&&&&A_{\alpha p}^p&&&&&\cr}\cr
 &B^{tr}_\alphadot=\pmatrix{
 B_{\alphadot 1}^1 &&&&&&&&&&\cr
 &B_{\alphadot 2}^2 &&&&&&&&&\cr
 && B_{\alphadot 3}^3 &&&&&&&&\cr
 &&&. &&&&&&&\cr
 &&&&.&&&&&&\cr
 &&&&&B_{\alphadot p}^p&&&&&}\cr
 &\sum_\alpha |A_{\alpha a}^a|^2 - \sum_\alphadot |B_{\alphadot
 a}^a|^2 =0 \qquad \forall a
 }}
At a generic point along this moduli space of vacua the gauge
group is broken to $SU(M)\times U(1)^{p-1}$.  The moduli space can
be characterized by $p$ sets of coordinates
$z_{\alpha\alphadot}^a= A_{\alpha a}^a B_{\alphadot a}^a$ with
$\det_{\alpha\alphadot} z_{\alpha\alphadot}^a =0$ up to
permutations over the index $a$. This is a symmetric product of
$p$ copies of the (singular) conifold $\CC_0$
 \eqn\clasmof{Sym_p(\CC_0)}
In addition to the $SU(M)$ gauge multiplets, the low energy
spectrum at a generic point includes $3p$ chiral multiplets and
$p-1$ vector multiplets.

If $U(1)_{baryon}$ is gauged, we have in addition to the $SU(M)$
part, $p$ multiplets of $\CN=4$. If we also add a Fayet-Iliopoulos
term $\xi$ for the $U(1)_{baryon}$, then supersymmetry is broken in all
these vacua. To leading order in the $U(1)_{baryon}$ gauge
coupling $g$, the vacua are given by \claflaf\ and the vacuum
energy is simply
 \eqn\clasxi{V = { g^2 \xi^2\over 2}\ .}

Even though this section is devoted mainly to the classical physics of
the model, we would like to make some comments about the
semi-classical low energy dynamics of the unbroken $SU(M)$ gauge
theory. It is a subgroup of the original $SU(N_1=M+p)$ and its
instanton factor is
 \eqn\lowenL{\Lambda_1^{3M+p} h^p \sim \Lambda(M,P,l=0)^{3M}}
This expression agrees with our general expression for the scale
of the unbroken group \scalel.  Equation \lowenL\ follows from
matching factors from the Higgsing $SU(M+p)\to SU(M)$ and from the
fields which get masses proportional to $h$.   It is consistent
with the (anomalous) symmetries of the problem \basissy.

The nonperturbative dynamics of the low energy gauge group leads
to gluino condensation and to a superpotential
 \eqn\WLowc{W= M \Lambda(M,P,l=0)^{1\over M} = M (\Lambda_1^{3M+p}
 h ^p)^{1\over M}}
as in \suplg.  The $M$ branches of \WLowc\ lead to $M$ vacua for
each point in \clasmof.  This is the origin of the sum over $r$ in the
quantum moduli space \quanm.  Note that since \lowenL\ and \WLowc\
are independent of the moduli in \claflaf, the
flat directions are not lifted. Instead, as we will see below, the moduli space
\clasmof\ is deformed.  Even though the superpotential \WLowc\
does not lead to a potential, it is important for determining the
tensions of domain walls connecting the different $M$ vacua
\refs{\DvaliXE\Kovner-\WittenEP}
(see section 11).

Let us examine the limit where $A_{\alpha p}^p$ and $B_{\alphadot
p}^p$ are much bigger than all other entries in \claflaf.  Then,
the low energy theory is an $SU(N_1-1=M+p-1) \times SU(N_2-1=p-1)
\times U(1)$ gauge theory with matter fields $A$ and $B$, as in
\matter, and three neutral chiral multiplets, whose vevs are
$A_{\alpha p}^pB_{\alphadot p}^p$.  This is the same as the
original theory except that $p$ is reduced to $p-1$, we have three
more neutral chiral fields and a $U(1)$ factor.  The instanton
factors of $SU(N_1-1=M+p-1) \times SU(N_2-1=p-1)$ are related to
the original ones \instfac\ as
 \eqn\scalestep{\hat\Lambda_1^{3M+p-1}\sim \Lambda_1^{3M+p} h
 \qquad ; \qquad \hat\Lambda_2^{p-2M-1} \sim \Lambda_2^{p-2M} h}
Note that these relations are independent of the vev of $A_{\alpha
p}^pB_{\alphadot p}^p$. Iterating this equation $p$ times leads to
\lowenL. The $U(1)$ factor originates from the two gauge groups
$U(1) \subset SU(N_1) \times SU(N_2)$. It is important that the
massless components of $A$ and $B$ are charged under this $U(1)$.
Therefore, in this low energy $SU(N_1-1) \times SU(N_2-1)\times
U(1)$ theory, the $U(1)_{baryon}$ is gauged.

\subsec{Baryonic flat directions for $\tilde p=0$, i.e.\
$N_1=(k+1)M$, $N_2=kM$}

For $\tilde p=0$ we find, in addition to the mesonic branch
\clasmof, two baryonic flat directions:
  \eqn\claflafg{\eqalign{
 &A_{\alpha=1}=C\pmatrix{
    \sqrt k    &0&0&.&0&0\cr
 0& \sqrt{k-1}   &0&.&0&0\cr
 0&0& \sqrt{k-2}   &.&0&0\cr
 .&.&.&.             &.&.\cr
 0&0&0&.& 1            &0\cr}\cr
 &A_{\alpha=2}=C\pmatrix{
 0    &1          &0&.&0&0\cr
 0&0& \sqrt 2       &.&0&0\cr
 0&0&0& \sqrt 3       &.&0\cr
 .&.&.&.&.              &.\cr
 0&0&0&0&.&            \sqrt k  \cr}\cr
 &\qquad\cr
 &B_{\alphadot=1}=0\cr
 &B_{\alphadot=2}=0
 }}
and another branch with $A \longleftrightarrow B$.  Here $C$ is an
arbitrary complex number and each entry in the matrices is an
$M\times M$ unit matrix.  The real constant $ {\cal U}  $ in
\cdterm\ is given by $ {\cal U} =k(k+1)M|C|^2$ for the solution
\claflafg\ and by $ {\cal U} =-k(k+1)M|C|^2$ for the solution with
$A \longleftrightarrow B$. The gauge invariant operator which is
nonzero along these branches is the baryon $(A_1A_2)^{k(k+1)M/2}$
(with appropriate contraction of the indices) or the anti-baryon
$(B_1B_2)^{k(k+1)M/2}$. Each of these branches is one complex
dimensional and is labelled by $C$ (more precisely, by
$C^{k(k+1)M}$), and they touch each other at the origin, $C=0$. We
refer to these branches as baryonic because, in contrast to the
mesonic branch \claflaf, here $U(1)_{baryon}$ is broken for
nonzero $C$. In fact, as we will see, these two branches are
joined in the quantum theory into a single smooth branch.

The low energy theory along each baryonic branch includes a chiral
superfield $C$ and an unbroken $SU(M)$.  Unlike the unbroken
$SU(M)$ on the mesonic branch, here $SU(M) \subset SU(M)_{N_1}
\times SU(M)_{N_2}$ where $ SU(M)_{N_1}\subset SU(M)^{k+1} \subset
SU(N_1=(k+1)M)$ and $ SU(M)_{N_2}\subset SU(M)^k \subset
SU(N_2=kM)$; i.e.\ the index of the embedding in $SU(N_1)$ is
$k+1$ and the index of the embedding in $SU(N_2)$ is $k$.  The
$SU(M)$ instanton factor is not given by \lowenL, but by
 \eqn\lowenLB{ \Lambda_1^{(k+1)(k+3)M}
 \Lambda_2^{k(k-2)M}h ^{2k(k+1)M} \sim \Lambda(M,p=kM,l=k)^{3M}}
Here we used the index of the embedding in the two groups and
the contribution from the matter fields which acquired mass from
the superpotential $W_0$.  Again, note that this relation is
independent of the modulus $C$.  This agrees with our general
expression for the scale of the unbroken group \scalel. Again,
gluino condensation leads to a superpotential
$M\Lambda(M,p=kM,l=k)^{1\over M}$ and hence to $M$ vacua.  Since
\lowenLB\ is independent of the modulus $C$, the flat direction is
not lifted.

If $U(1)_{baryon}$ is gauged, $C$ must vanish, and it seems that
the theory is at the origin of field space.  We will see below
that this is not true in the quantum theory where in this case the
theory has isolated vacua with broken $U(1)_{baryon}$.  One way to
see that is to add a Fayet-Iliopoulos  term $\xi$ for
$U(1)_{baryon}$.  Then, depending on the sign of $\xi$ either $A$
or $B$ is nonzero as in \claflafg\ and $C$ is fixed in terms of
$\xi$.  More explicitly, for positive $\xi$ we have the solution
\claflafg\ with
 \eqn\Cintermsxi{\xi= {\cal U}=k(k+1)M|C|^2\ .}
Since $C\not=0$, the gauged $U(1)_{baryon}$ symmetry is Higgsed,
the phase of $C$ acquires a mass, and the vacuum is isolated.

Now we return to the case where $U(1)_{baryon}$ is not gauged.
For $\tilde p \not=0$ the only flat directions are the mesonic
ones, \claflaf. One way to understand it is by trying to reduce $p$
using expectation values as in \claflaf\ down to $p=k M$, and
looking for a baryonic solution similar to \claflafg\ using the
massless fields. More explicitly, consider \claflaf\ with
$A_{\alpha a}^a = B_{\alphadot a}^a=0$ for $a=1,...,kM$.  The low
energy theory is an $SU((k+1)M) \times SU(kM) \times U(1)^{\tilde
p}$ gauge theory with charged fields as in \matter, and neutral
chiral fields taking values in $\tilde p$ copies of $\CC_0$. The
charged matter in this theory is very similar to the that of the
$\tilde p=0$ theory which leads to \claflafg, except for one important
difference. The charged chiral fields are charged under one linear
combination of $U(1)^{\tilde p}$. As we commented above, in this
case $C$ in \claflafg\ must vanish, so this does not lead to new
flat directions.  Below we will see how this fact is modified in
the quantum theory.

\newsec{$SU(N_1=M+p)\times SU(N_2=p)$ with $p=0,...,M-1$}

Here we discuss the quantum theory for small values
of $p$.
As we have already commented, in the quantum theory the
$SU(M)$ that remains unbroken
along the flat directions becomes strong and leads to $M$
vacua. Hence, the previous answer for the moduli space appears $M$
times. Yet, this is not the whole story.

In order to examine it in more detail we first ignore the tree
level superpotential, $W_0$, and the weaker of the two gauge groups,
$SU(p)$.  Since the first group, $SU(N_1=M+p)$, has fewer flavors
than colors ($2N_2=2p<N_1$),  its flat directions are
characterized by the meson fields
 \eqn\mesonf{\CM_{\alpha \alphadot a}^b = A_{\alpha i}^a
 B_{\alphadot i}^b\ .}
At a generic point along these flat directions, the unbroken gauge
symmetry is $SU(N_1-2N_2=M-p)\subset SU(N_1=M+p)$.  This gauge
theory confines and its dynamics generates the superpotential
\refs{\AffleckMK,\AffleckXZ}
 \eqn\weffoa{W_{dyn}=(N_1-2N_2) \left({\Lambda_1^{3N_1 - 2 N_2}
 \over \det_{\alpha\alphadot a b} \CM}\right)^{1\over
 N_1-2N_2}=(M-p)  \left({\Lambda_1^{3M +p } \over
 \det_{\alpha\alphadot a b} \CM}\right)^{1\over M-p}}
The fractional power in the superpotential is associated with the
$M-p$ vacua of $SU(M-p)$. Therefore, we should study the dynamical
superpotential as a function on an $M-p$ fold cover of the space
of $\CM$.

So far our description has neglected the tree level superpotential
$W_0$ and the D-term equations of the second gauge group,
$SU(N_2=p)$. Therefore, $\CM$ is generic. On the other hand, the
typical points on the classical flat directions \claflaf\ have
non-generic $\CM$, such that $\det_{\alpha\alphadot a b} \CM=0$.
For generic $\CM$ the $SU(N_1)$ gauge symmetry is broken to
$SU(N_1-2N_2=M-p)$, while along the classical flat directions of
the full theory \claflaf\ the $SU(N_1)$ gauge group is broken to
$SU(M)$.  We will now show that restoring $W_0$ and the
$SU(N_2=p)$ interactions restricts us to a subspace of $\CM$. This
subspace is not the same as the classical moduli space \claflaf;
it is a deformation of it.

Consider a generic point in $\CM$.  After the unbroken
$SU(N_1-2N_2=M-p)$ confines and leads to \weffoa, the low energy
theory is an $SU(N_2=p)$ gauge theory with neutral matter fields
$\CM$ of \mesonf\ (four adjoints and four singlets), and the
superpotential
 \eqn\weffo{W_{eff}=W_0+W_{dyn}}
This theory is IR free and can be analyzed easily.

Solving $\partial_\CM W_{eff}=0$ we find
 \eqn\stawf{\eqalign{
 &h^p\det_{\alpha\alphadot a b} \CM \sim \left(\Lambda_1^{3M+p} h^p
 \right)^{p\over M}\cr
 &h \Tr_a \det_{\alpha\dot\alpha} \CM_{\alpha\alphadot} \sim \left( h^p
 \Lambda_1^{3M+p}\right)^{1\over M}
 }}
Although classically $\det \CM=0$, in the quantum theory
this is deformed by nonperturbative effects.  As a check, note
that in the $\Lambda_1\to 0$ limit the determinant vanishes.
Related to that is the appearance of a fractional power $1\over M$. It reflects
the fact that $SU(N_1)$ is broken to $SU(M)$ with no charged
matter with scale
 \eqn\aasc{ \Lambda_1^{3M+p} h^p \sim \Lambda(M,P,l=0)^{3M}\ ,}
which agrees with our general expression for the unbroken group
and its instanton factor \scalel\ and \lowenL. The strong dynamics
of this $SU(M)$ theory leads to $M$ vacua.

Finally, in addition to \stawf\ we also need to impose the D-term
equations of $SU(N_2)$. The conclusion is that the moduli space
can be parameterized by the $4p$ numbers $\CM_{\alpha\alphadot
a}^a$ with $a=1,...,p$, $\alpha,\alphadot=1,2$ subject to the
constraints
 \eqn\cons{h \det_{\alpha\alphadot} \CM_{\alpha\alphadot a}^a
 =\epsilon_{M,p}(r,l=0) \sim \left( h^p
 \Lambda_1^{3M+p}\right)^{1\over M} \qquad \forall a}
i.e.\ they are on the deformed conifold $\CC_{r,l=0}$ \defconz\
with $\epsilon$ as in \epsilonl. Of course, we should mod out this
space by the permutation of the $p$ points, and therefore the
moduli space is $\oplus_r Sym_p(\CC_{r,l=0})$.

Classically, the answer \clasmof\ is related to the conifold
$\CC_0$. The nonperturbative effects associated with a power
of $\Lambda_1$ \cons\ deform it to the deformed conifold
$\CC_{r,l}$. More precisely, we work on an $M-p$ fold cover of the
space of $\CM$ and find $M$ different solutions of \stawf.

The final answer is
 \eqn\modulio{ \oplus_{r=1}^M Sym_p(\CC_{r,l=0})}
At a generic point in this space the $SU(p)$ is broken to
$U(1)^{p-1}$.  The $3p$ massless chiral multiplets describe the
positions of $p$ D3-branes in the deformed conifold
$\CC_{r,l}$.\foot{If $U(1)_{baryon}$ is gauged, the low energy
modes combine into $p$ vector multiplets of $\CN=4$.}

The answer \modulio\ can be interpreted as $p$ pairs of
D5-anti-D5-branes moving from the tip to the bulk of the
deformed conifold $\CC_{r,l}$ as $p$ D3-branes.  The different
phases of $\epsilon$ are different fluxes through the cycle of the
deformed conifold.
To reach this conclusion we needed the
dynamical superpotential in \weffo\ -- the tree level theory based
on the superpotential $W_0$ does not lead to this answer.

To summarize, the answer \modulio\ differs from the classical
answer derived from \claflaf\ in two ways.  First, we have $M$
branches originating from the $SU(M)$ gauge dynamics. Second,
the argument of the symmetric product $\CC_0$ is deformed to
$\CC_{r,l=0}$. As expected, far from the origin of the moduli space
the classical analysis, together with the existence of $M$ vacua in
the strongly coupled low energy theory, gives a good approximation
to the quantum answer.

\newsec{$SU(N_1=2M)\times SU(N_2=M)$; i.e.\ $p=M$}

In this case the $SU(N_1)$ theory has equal numbers of flavors and
colors. Therefore, in addition to the mesons $\CM_{\alpha
\alphadot a}^b$ of \mesonf\ which are four adjoints and four
singlets of $SU(N_2)$, the theory has baryons
 \eqn\baryonsf{\eqalign{
 &\CA = \epsilon^{i_1...i_{N_1}}A_{\alpha_1 i_1}^{a_1} ...
 A_{\alpha_{N_1} i_{N_1}}^{a_{N_1}}\ ,\cr
 &\CB = \epsilon_{i_1...i_{N_1}}B_{\alphadot_1 a_1}^{i_1}
 ... B_{\alphadot_{N_1} a_{N_1}}^{i_{N_1}} \ .
 }}
Bose statistics shows that these are singlets of the non-Abelian
symmetries in the problem \SeibergBZ, and in particular of
$SU(N_2)$ and $SU(2)\times SU(2)$
\refs{\KlebanovHB,\AharonyPP,\GubserQJ}. The fields $\CM$, $\CA$
and $ \CB$ are not independent.  They are subject to the
constraint \SeibergBZ\ $\det_{\alpha\alphadot a b} \CM - \CA  \CB
=\Lambda_1^{2N_1}$ where we have already taken the nonperturbative
quantum corrections into account.  This can be summarized by an
effective superpotential \SeibergBZ
 \eqn\wefft{W_{eff}=W_0+L \left(\det_{\alpha\alphadot a b} \CM -
 \CA  \CB -\Lambda_1^{2N_1}\right)}
where $L$ is a Lagrange multiplier.\foot{It is important to clarify
a common misconception.  The superpotential \wefft\ is not a low
energy superpotential.  It includes fields which are massive
everywhere on the moduli space, in particular, the Lagrange
multiplier $L$ and one component of $\CM$, $\CA$ and $\CB$ are
always massive. These fields are not associated with massive
particles in the spectrum. Instead, they should be interpreted as
auxiliary fields in the low energy theory.} The low energy theory
is based on an $SU(N_2)$ gauge theory with these massless fields
and the superpotential \wefft. It is IR free and is easily
analyzed.

The moduli space has two branches which are related to the
branches of the classical moduli space.

The mesonic branch has $l=0$.  It is characterized by $\CA= \CB=0$
and $\CM$ constrained by $\det_{\alpha\alphadot a b} \CM
=\Lambda_1^{2N_1}$. It is also subject to the $SU(N_2)$ D-term
equations and stationarity of $W_0$. This leads to the moduli
space $\oplus_r Sym_{p=M}(\CC_{r,l=0})$ with the deformation
parameter $\epsilon_{M,p=M}(r,l=0)$ of \epsilonl, as in \cons. At
a generic point on this branch the theory has $M-1$ vector
multiplets and $M$ chiral multiplets corresponding to the motion
of the $M$ D3-branes on $\CC_{r,l=0}$.\foot{If we gauge the
$U(1)_{baryon}$, we find $p=M$ multiplets of $\CN=4$.}

The second branch is baryonic.  It has $\CM=0$ and $\CA  \CB =
\Lambda_1^{2N_1}$.  As discussed above, the low energy theory
includes a pure gauge $SU(M)$ sector which leads to $M$ vacua.
Each of the $M$ components of the baryonic branch is one complex
dimensional.  Note that the two classical branches with $\CA
\CB=0$ are combined into a single smooth branch with $\CA \CB =
\Lambda_1^{2N_1}$.

If we gauge $U(1)_{baryon}$, the baryonic branch becomes zero
dimensional.  It has $M$ discrete points -- a point for each
component of the baryonic branch labelled by $r=1,...,M$. Each of
them has a gap because the $U(1)_{baryon}$ gauge symmetry is
Higgsed \GubserQJ. The existence of such points is completely
quantum mechanical.  They originate from the singularity at the
origin of the classical moduli space.  In the quantum theory this
singularity leads to several vacua.  The mesonic branch which
classically touches the origin is deformed.  In addition, the
origin leads to these isolated vacua where $U(1)_{baryon}$ is
broken.  We now turn on a nonzero Fayet-Iliopoulos term $\xi$.  As
above, supersymmetry is broken in the mesonic branch.  The $M$
isolated vacua remain supersymmetric, but their position in field
space changes. As $\xi \to \infty$ these supersymmetric vacua move
to large field strength where they are continuously connected to
the classical vacua discussed around \Cintermsxi.

In different regions of the baryonic branch the low energy $SU(M)$
gauge theory and its scale can be understood differently.  First,
far out along the baryonic branch \claflafg\ we can use the
general result \lowenLB.  Specializing to $k=1$, it is
 \eqn\lowenLBa{ \Lambda_1^{8M}
 \Lambda_2^{-M}h ^{4M} \sim \Lambda(M,p=M,l=1)^{3M}}
Near the origin, where $SU(N_1=2M)$ is strongly coupled, this
calculation is not valid. Instead, we can find $ \Lambda(M,p=M,l=1)$ as follows.
Above the scale $\Lambda_1$ the second gauge group $SU(N_2=M)$ has
$4M$ fundamental flavors, and its instanton factor is
$\Lambda_2^{-M}$.  Below $\Lambda_1$ the fundamental flavors are
confined and they are replaced by four adjoints $\CM$.  The
instanton factor $\Lambda_2^{-M}$ does not change.  Then the
adjoints get mass $h \Lambda_1^2$ and the resulting instanton
factor is $\Lambda_2^{-M}(h \Lambda_1^2)^{4M}$.  Miraculously,
this different calculation which is based on different physics
agrees with \lowenLBa, and also with the general expression
\scalel.

We have already stated that the expectation values of the mesons
are interpreted as D3-branes in the bulk of the conifold.  We
would like to interpret the baryonic branch as describing {\it a
BPS bound state at threshold of $2M$ D5-branes and $M$
anti-D5-branes.}  If $U(1)_{baryon}$ is not gauged, the baryon
number is broken by the bound state and it is part of a one
complex dimensional branch of the moduli space.

\newsec{$SU(N_1=2M+1)\times SU(N_2=M+1)$; i.e.\ $p=M+1$}

We now consider the case $p=M+1$.  Here the $SU(N_1)$ gauge theory
interacts with $2N_2=2M+2=N_1+1$ flavors; i.e.\ it is the number
of colors plus one. Its dynamics leads to the mesons $\CM$ of
\mesonf\ and baryons $\CA_{\alpha a}$, $\CB_{\alphadot}^a$ similar to \baryonsf\
in that they have opposite
charges under the baryon number symmetry $U(1)_{baryon}$.
However,
now they are in the anti-fundamental and the fundamental of
$SU(N_2)$ respectively \SeibergBZ. Each of them transforms under one of
the two factors in $SU(2) \times SU(2)$.

The low energy theory includes these mesons, baryons and the gauge
fields of $SU(N_2)$ with a superpotential \SeibergBZ\
 \eqn\weffth{W_{eff}=W_0-{1\over \Lambda_1^{3N_1-2N_2}}\left(
 \det_{\alpha\alphadot a b} \CM - \CA \CM  \CB \right)}

The moduli space is easily determined and  again has two
branches.  On one of them, $\CA= \CB=0$ and $\CM$ satisfies \stawf\
as well as the $SU(p)$ D-term equations. This leads to $\oplus_r
Sym_{p=M+1}(\CC_{r,l=0})$ with the deformation parameter
$\epsilon$ as in \epsilonl\ and \cons
 \eqn\defpnpoa{\epsilon_{M,p=M+1}(r,l=0) \sim
 \left(h^{M+1} \Lambda_1^{4M+1}\right)^{1\over M}}

On the other branch  $\CM$ is massive and should be integrated
out. Clearly
 \eqn\cmvev{\CM_{\alpha \alphadot a}^b \sim  {1\over h
 \Lambda_1^{4M+1} } \epsilon_{\alpha \alphadot }
 \epsilon_{\beta \dot \beta}  \CB_{\dot \beta}^b\CA_{\beta
 a}}
($\CO((\CB \CA)^2)$ corrections from $\det \CM$ in \weffth\ are
not present because $\CM$ of \cmvev\ has low rank.)  This leads to
a superpotential of the form
 \eqn\nposup{{1\over h \Lambda_1^{8M+2}}
 \det_{\alpha\alphadot} (\CA_{\alpha a} \CB_\alphadot^a) }
In addition, the low energy theory has an $SU(p=M+1)$ gauge theory
under which $\CA$ and $ \CB$ transform.

This theory is similar to the one discussed in the section 5 about
$p<M$, if we use
 \eqn\pMchange{\hat M = M \qquad ; \qquad \hat p = 1}
i.e. it has gauge group $SU(\hat N_1=\hat M + \hat p = M+1)$ and
$\hat N_2=1$.  Therefore, we can borrow the result of the analysis
there.  In order to do that we have to bring it to a canonical
form and relate its parameters to the parameters of the ``hat
theory.''

First, the ``quarks'' $\CA_\alpha$ and $ \CB_\alphadot$ do not
have their canonical dimensions. Therefore we define the canonical
fields $\hat \CA_\alpha = \CA_\alpha/\Lambda_1^{2M}$ and $\hat
\CB_\alphadot = \CB_\alphadot/\Lambda_1^{2M}$.  Their quartic
coupling is
 \eqn\hhat{\hat h = {1\over h \Lambda_1^2}}
The scale of the $SU(N_2=M+1)$ is also modified. The instanton
factor of the high energy theory is
$\Lambda_2^{p-2M}=\Lambda_2^{1-M}$. Below the scale $\Lambda_1$
the $SU(N_2)$ gauge theory has two fundamental flavors $\CA$ and
$\CB$ and four adjoints $\CM$. Therefore, its instanton factor is
$ \Lambda_L^{3M+3-2-4(M+1)}= \Lambda_L^{-3-M}\sim \Lambda_2^{1-M}
\Lambda_1^{-4}$.  Then, at lower energies the four adjoints get a
mass of order $h \Lambda_1^2$ and the instanton factor of the
$SU(M+1)$ gauge theory is
 \eqn\scalehat{ \hat \Lambda_1^{3M+1}\sim \Lambda_L ^{-3-M}(h
 \Lambda_1^2)^{4(M+1)} \sim
 \Lambda_2^{1-M} h ^{4(M+1)} \Lambda_1^{8M+4 }}

Now it is straightforward to use the results from section 5. The
moduli space is $M$ copies of $ \CC_{r,l}$ with
 \eqn\tempdef{\hat h \det_{\alpha\alphadot} \hat \CA_{\alpha a}
 \hat \CB_\alphadot^a \sim  \hat \epsilon _{\hat M=M, \hat
 p=1}(r,l=0) \sim  (\hat h \hat \Lambda_1^{3M+1})^{1\over  M}}
Expressing it in terms of $\CA$ and $ \CB$ and finally in terms of
the eigenvalues of $\sqrt h \CM$ we find the deformed conifold
with deformation parameter
 \eqn\defpnpo{\epsilon_{M,p=M+1}(r,l=1) \sim
 \left(h^{4M+3} \Lambda_1^{8M+2}\Lambda_2^{1-M}\right)^{1\over M}}
Note the combination
 \eqn\lowlapo{h^{4M+3} \Lambda_1^{8M+2}\Lambda_2^{1-M}
  \sim \Lambda(M,p=M+1,l=1)^{3M}}
which can be interpreted as the instanton factor of the low energy
$SU(M)$ theory.  The expressions \defpnpo\lowlapo\ are in
agreement with our general expression \scalel\epsilonl. Finally,
the low energy spectrum has three chiral superfields.\foot{If we
gauge $U(1)_{baryon}$, the low energy spectrum forms a $\CN=4$
supermultiplet.}

We would like to make a few comments about this branch. First,
even though $\CA$ and $ \CB$ carry baryon number, their
expectation values do not mean that $U(1)_{baryon}$ is broken.
Instead, $U(1)_{baryon}$ combined with a broken gauge symmetry is
unbroken. One way to see that is to note that $\CA$ and $\CB$ are
not gauge invariant.  The order parameter along the moduli space
is $ \CA_{\alpha a} \CB_\alphadot ^a$ which is $U(1)_{baryon} $
neutral. Equivalently, the relation \cmvev\ along the moduli space
shows that baryon number is not broken, and therefore we cannot
refer to this branch as a baryonic branch.

Second, we should clarify the relation to the classical analysis
in section 4.  Equation \cmvev\ shows that in the classical limit
$\Lambda_1 \to 0$ the order parameters $\CA$ and $ \CB$ must
vanish in order to have finite $\CM$.  This shows that the
semiclassical limit of these vacua is given by $\CM$ with rank
one.  We have discussed these vacua when we looked for classical
baryonic branches of this theory.  We separated one eigenvalue of
the meson field in order to have a low energy theory with $p=M$
which has a baryonic branch.  There we saw that the low energy
theory has a gauged baryon number, and therefore there is no
classical baryonic branch.  As we stated above, even when
$U(1)_{baryon}$ is gauged, the quantum $p=M$ theory has isolated
vacua in which $U(1)_{baryon}$ is Higgsed. The branch we have been
discussing here in the $p=M+1$ theory is associated with such a
mesonic eigenvalue and such an isolated vacuum of the low energy
theory.  In terms of the branes on the conifold, it is a single
D3-brane in the bulk of $\CC_{r,l=1}$ and a bound state at
threshold of the other branes.

Finally, we would like to point out a subtlety in integrating out
$\CM$.  There is no problem with doing it along the $l=1$ branch,
as in \cmvev, \nposup. (Note that this integration out obscures
the semi-classical limit.) However, such integration out is
impossible along the mesonic branch with $l=0$, because there the
field $\CM$ has massless components.  We should stress though that
\weffth\ is valid on both branches.

In conclusion, our moduli space is
 \eqn\modmpoc{\oplus_{r=1}^M \left[Sym_{M+1}(\CC_{r,l=0}) \oplus
 \CC_{r,l=1} \right]}
The first term is interpreted as $p=M+1$ D3-branes on the deformed
conifold.  On the second branch there is a BPS bound state at
threshold of $2M$ D5-branes and $M$ anti-D5-branes while the
remaining D3-brane is free to move on the deformed conifold
$\CC_{r,l=1}$.

\newsec{$SU(N_1=M+p)\times SU(N_2=p)$ with $M+1<p$}

The previous case of $p=M+1$ has almost all the elements the we
need for the general case of larger $p$.

The first branch of the moduli space is a deformation of \claflaf.
As above, we first ignore the $SU(N_2)$ dynamics and the tree
level superpotential ($h=0$).  The $SU(N_1)$ dynamics has many
more flavors than colors, and therefore the rank of the meson $\CM$
is constrained.  Nevertheless, some of the dynamics can be
recovered by considering $\CM$ of maximal rank $2p$ and the
superpotential \weffoa
 \eqn\weffoaa{W_{dyn}=(N_1-2N_2) \left({\Lambda_1^{3N_1 - 2 N_2}
 \over \det_{\alpha\alphadot a b} \CM}\right)^{1\over
 N_1-2N_2}=(M-p)  \left({\Lambda_1^{3M +p } \over
 \det_{\alpha\alphadot a b} \CM}\right)^{1\over M-p}}
As in all our cases, solving $\partial_\CM W_{eff}=0$ we find
\stawf
 \eqn\stawfa{\eqalign{
 &h^p\det_{\alpha\alphadot a b} \CM \sim \left(\Lambda_1^{3M+p} h^p
 \right)^{p\over M}\cr
 &h \Tr_a \det_{\alpha\dot\alpha} \CM_{\alpha\alphadot} \sim \left( h^p
 \Lambda_1^{3M+p}\right)^{1\over M}
 }}
This is the mesonic $l=0$ branch $\oplus_{r=1}^M
Sym_{p}(\CC_{r,l=0})$ with deformation parameter
$\epsilon_{M,p}(r,l=0)$ and low energy $SU(M)$ dynamics with scale
$\Lambda(M,p,l=0)$.  This branch describes $p$ D3-branes on the
deformed conifold.

The other branches involve the strong coupling dynamics.  The
easiest way to find them is to dualize the $SU(N_1)$ theory as in
\SeibergPQ . This leads to an $SU(2N_2-N_1=p-M)$ gauge theory with
dual quarks $\hat A$ and $\hat B$, and the meson $\CM$.  Restoring
the second gauge group $SU(N_2)$ we have a theory which is similar
to the original one except that $p\to p-M$ and the superpotential
is
 \eqn\dualsup{{1\over \mu}\hat A \CM \hat B + h \CM \CM}
(the parameter $\mu$ and its role are explained in
\IntriligatorAU.)  For generic $\CM$ the dual quarks $\hat A$ and
$\hat B$ acquire a mass and then the dual gauge group
$SU(2N_2-N_1=p-M)$ can be integrated out leading back to \weffoaa.
This way we recover the previously discussed mesonic branch with
$l=0$. However, this theory also has another branch which is not
obvious in the original degrees of freedom.

We can integrate out $\CM$ in \dualsup\
 \eqn\intM{\CM \sim {1\over \mu h} \hat B \hat A}
to find a theory which is similar to our original theory except
that $p \to p-M$. \foot{As explained in section 7 about the
$p=M+1$ theory, after we have done this we can no longer recover
the $l=0$ branch.} This theory has several branches. One of them
is a mesonic branch $\oplus_{r=1}^M Sym_{p-M}(\CC)$, whose
deformation parameter will be determined shortly. This branch
arises similarly to the $l=1$ branch for $p=M+1$ where the role of
the dual quarks was played by $\CA$ and $ \CB$. The
$U(1)_{baryon}$ is again unbroken, and the classical limit of this
branch coincides with a subspace of the $l=0$ mesonic branch. We
interpret it, as there, in terms of a bound state at threshold of
$2M$ D5-branes and $M$ anti-D5-branes. In addition, we find $p-M$
D3-branes moving on $\CC_{r,l=1}$.

Now let us analyze the parameters of the low energy theory
generalizing the discussion for $p=M+1$ in section 7. For
simplicity, we set the parameter $\mu$ in \dualsup\ equal to the
scale of the group which is being dualized, $\Lambda_1$. First, we
write the gauge theory as
 \eqn\newg{SU(\hat N_1=p) \times SU(\hat N_2=p-M)}
to agree with our general notation with $N_1 \ge N_2$.  Therefore
we have
 \eqn\mphat {\hat M=M \qquad ; \qquad \hat p =p-M}
Next, it is clear that the quartic coupling is
 \eqn\hhatg{\hat h  \sim {1 \over h \Lambda_1^2}}
The instanton factor of the second group $SU(\hat N_2)$ is related
to its dual $SU(N_1)$ using $\Lambda_1^{3M+p} \hat
\Lambda_2^{p-3M} \sim \mu^{2p}=\Lambda_1^{2p}$ and therefore
 \eqn\hatlambdat{\hat \Lambda_2 = \Lambda_1}
The instanton factor of the microscopic $SU(N_2=p)$ is
$\Lambda_2^{p-2M}$. After the duality this theory has $2(p-M)$
fundamental flavors and four adjoints. Therefore, its instanton
factor is $\Lambda_L^{3p-2(p-M)-4p}= \Lambda_L^{2M-3p}\sim
\Lambda_2^{p-2M} \Lambda_1^{4(M-p)}$.  After the adjoints get a
mass of order $h \Lambda_1^2$, the scale of the $SU(\hat N_1=N_2)$
theory is
 \eqn\scaletwohg{\hat\Lambda_1^{p+2M} \sim \Lambda_L^{2M-3p} (h
 \Lambda_1^2)^{4p} \sim h^{4p}
 \Lambda_1^{4(M+p)}\Lambda_2^{p-2M}}
Now, we can use our earlier result about the $l=0$ branch of this
low energy theory. Using \mphat-\scaletwohg\ we can express the
results in terms of the original microscopic parameters
 \eqn\hatlambdalone{\eqalign{
 \hat \Lambda(\hat M,\hat p, l=0)^{3\hat M} \sim & \hat h^{\hat p}
 \hat \Lambda_1^{3 \hat M + \hat p} \sim (h \Lambda_1^2)^{M-p}
 (h^{4p} \Lambda_1^{4(M+p)}\Lambda_2^{p-2M}) \cr
  = & h^{M+3p} \Lambda_1^{2(3M+p)}\Lambda_2^{p-2M} \sim
  \Lambda( M, p, l=1)^{3M}}}
exactly as in \scalel.

Using \hatlambdalone\ we can find the deformation parameter for
$\hat A_{a i}\hat B^{b i}$ and finally for the eigenvalues of the
meson $\sqrt h \CM$ (use \intM\ with $\mu = \Lambda_1$)
  \eqn\epsilonlo{
  \epsilon_{ M, p}(r,l=1) \sim \,  {h\over (h \Lambda_1)^2 \hat h}
  \hat \epsilon_{\hat M,\hat p}(r,l=0) \sim\,
 \left(\hat \Lambda(\hat M,\hat p,l=0)^{3M} \right)^{1\over M}
 \sim \, \left(\Lambda(M,p,l=1)^{3M}\right) ^{1\over M}}
which agrees with the general expression in \epsilonl.

Now it is clear how we can continue dualizing this way to find our
final answer for the moduli space
 \eqn\quanmf{\oplus_{r=1}^M \oplus _{l=0}^k
 Sym_{p-lM}(\CC_{r,l})}
with
 \eqn\epsilong{\eqalign{
 \epsilon_{M,p}(r,l)\sim\,  &
 \left(\Lambda(M,p,l)^{3M}\right)^{1\over M}\cr
 \Lambda(M,p,l)^{3M} \sim\, &h^{p+l(M+2p)}
 \Lambda_1^{(3M+p)(l+1)}\Lambda_2^{(p-2M)l}}}

After the duality, the integration out of the meson $\CM$ does not
handle correctly the branch of the moduli space of the largest
dimension ($l=0$), but the next branch with $l=1$ becomes
manifest.  This branch was interpreted as a bound state of
$2M$ D5-branes and $M$ anti-D5-branes near the tip of the
conifold. As we continue to dualize the story repeats itself and
we find more branches which correspond to bound states of $(l+1)M$
D5-branes and $lM$ anti-D5-branes.  The cascade stops when we use up
all the available D5-branes in this fashion.

\newsec{Consistency checks -- relations between theories and
branches}

By considering various boundaries of the moduli space we can find
relations between the different branches of different gauge
theories.  These relations provide nontrivial consistency checks
of our expressions \scalel\epsilonl.

First, consider the limit as one of the eigenvalues of $\CM$ is
moved to infinity.  In terms of the brane interpretation this
corresponds to removing a D3-brane from the system.  In the gauge
theory this has the effect of changing $p \to p-1$, and adding a
$U(1)$ factor.  In \scalestep\ we expressed the instanton factors
of this low energy $SU(N_1-1) \times SU(N_2-1)$ theory in terms of
the microscopic scales
 \eqn\scalestepa{\hat\Lambda_1^{3M+p-1}\sim \Lambda_1^{3M+p} h
 \qquad ; \qquad \hat\Lambda_2^{p-2M-1} \sim \Lambda_2^{p-2M} h}
Clearly, the parameter $h$ does not change, $\hat h = h$.  It is
easy to check that the transformations $p\to \hat p= p-1$ and
$\Lambda_{1,2} \to \hat \Lambda_{1,2}$ map
 \eqn\LoLtI{\eqalign{
 \hat L_1( M,\hat p)=\, & \hat h^{\hat p} \hat \Lambda_1^{3 M
 + \hat p} \sim \, L_1(M,p)\cr
 \hat L_2( M,\hat p)=\, & \hat h^{ M+ \hat p} \hat
 \Lambda_2^{\hat p - 2  M} \sim L_2(M,p)\cr
 \hat I( M, \hat p)=\, &\hat L_1( M,\hat p) \hat L_2(\hat
 M,\hat p) = I( M,  p)
 }}
and therefore our relations \scalel\epsilonl\ are mapped
consistently
 \eqn\epsilonla{\eqalign{
 \hat \Lambda(M,\hat p,l)^{3 M} \sim\,& \hat L_1( M,\hat p)
 \hat I( M, \hat p)^l \sim \, \Lambda(M,p,l)^{3M}\cr
 \hat \epsilon_{M,\hat p }(r,l)\sim \, & \left(\hat \Lambda(M,
 \hat p,l)^{3 M}\right)^{1\over M} \sim \, \epsilon_{M,p}(r,l)
 }}
As expected, the moduli space of the remaining branes is not
affected by removing a D3-brane.

Clearly, we can iterate this process as long as $\tilde p$ remains
positive.  As explained in the previous sections, if we continue
this way down to $\tilde p=0$ we do not find the one complex
dimensional baryonic branches.  Instead, we find them as zero
dimensional branches because effectively the baryon number
symmetry is now gauged.  As we continue down this road to smaller
values of $p$, we loose the branches with large values of $l$.
This is expected from the brane picture.  However, it is
important that the branches which are found have their correct
deformations and scales.

The process of reducing $p$ relates all $\Lambda(M,p,l)$ to
$\Lambda(M,p=lM,l)$.  Our second consistency check involves the
value of this scale.  When we discussed the $p=M$ theory in section 6,
we checked that $\Lambda(M,p=M, l=1)$ was obtained
correctly in two different regions of the moduli space \lowenLBa.
This discussion is easily generalized to higher $l$.  Far out
along the baryonic branch of the $\tilde p=0$ theory we have the
expression \lowenLB\ which is based on the nontrivial embedding of
$SU(M)$. Near the origin of the moduli space the same expression
is easily found using matching relations and the dual theory (we
do not give the details here).

These two checks are nontrivial tests of our expressions
\scalel\epsilonl\ and the brane interpretation.  They reinforce
the interpretation of the bound state being the same bound state
for any value of $p$ with the same $M$ and $l$.  Also, it is the
same state which is visible semiclassically along the baryonic
branch of the $\tilde p=0 $ theory.  Clearly, removing D3-branes
from the system should not affect the bound state.

Finally we comment that the duality transformations relate
different values of $l$:
 \eqn\lhatl{\hat l = l-1}
More explicitly, we now generalize \hatlambdalone\epsilonlo.
Every duality transformation maps the parameters as in
\newg-\scaletwohg
 \eqn\mphatg {\eqalign{
 \hat M=& M \cr
  \hat p =& p-M \cr
 \hat h  \sim \, & {1 \over h \Lambda_1^2}\cr
 \hat \Lambda_2 = & \Lambda_1\cr
 \hat\Lambda_1^{\hat p+2M} \sim\, &  h^{4p}
 \Lambda_1^{4(M+p)}\Lambda_2^{p-2M}
 }}
Therefore,
 \eqn\LoLtIse{\eqalign{
 \hat L_1( M,\hat p)=\, & \hat h^{\hat p} \hat \Lambda_1^{3 M
 + \hat p} \sim \, L_1(M,p) I (M,p)\cr
 \hat L_2( M,\hat p)=\, & \hat h^{ M+ \hat p} \hat
 \Lambda_2^{\hat p - 2  M} \sim {1\over L_1(M,p)}\cr
 \hat I( M, \hat p)=\, &\hat L_1( M,\hat p) \hat L_2(\hat
 M,\hat p) = I( M,  p)
 }}
 Then, using \lhatl\ we have
 \eqn\hatlambdaloneg{\eqalign{
 \hat \Lambda(\hat M,\hat p, \hat l)^{3\hat M} \sim & \, \hat L_1(
 M,\hat p) \hat I( M, \hat p)^{\hat l} \sim L_1(
 M, p)  I( M,  p)^{ l} \sim  \Lambda( M, p, l)^{3M} \cr
 \hat \epsilon_{\hat M,\hat p}(r,\hat l) \sim & \,  \left(\hat
 \Lambda(\hat M,\hat p,\hat l)^{3M} \right)^{1\over M} \sim
 \epsilon_{ M, p}(r, l)
 }}

This last consistency check, which is associated with a change of
$l$, has the following brane interpretation.  As we cascade down
the conifold we change the theory, $p \to p-M$, the parameters of
the theory \mphatg, and the branch $l \to l-1$.  This means that
the number of D3-branes which are free to move is unchanged but
the bound state includes fewer branes. Since this description is
valid only closer to the tip of the conifold, this means that this
bound state with smaller $l$ is physically smaller.  More
heuristically, we can think of these bound states as being large
atoms with many electrons.  As we cascade down to smaller $p$ and
smaller $l$ the space we look at is getting smaller and only
electrons in inner shells fit in the space and can be included in
the bound state.

The fact that the parameter $I(M,p)$ does not change under the
transformations \mphatg\ is consistent with our identification
$I(M,p)=e^{2\pi i\tau}$ in the type II theory.
In the weak string coupling limit $|I(M,p)|\ll
1$ the deformation parameter
\eqn\epsir{|\epsilon_{M,p}(r,l)| \sim
|\epsilon_{M,p}(r,l=0)| |I(M,p)|^{l/M}\sim
|\epsilon_{M,p}(r,l=0)| e^{-2\pi l/(g_s M)}
\ .}
This is in exact agreement with the dual string theory.\foot{The
following argument is due to J. Maldacena.}
We may embed the $SU(M+p)\times SU(p)$ gauge theory into a string
compactification as in \Giddings. Then, the 5-form flux conservation gives
the constraint $p = lM + N_{free}$. Here $N_{free}$ is the number of mobile
D3-branes, $M$ is the number of units of the RR 3-form flux though
the A-cycle,
and $l$ is the number of NS 3-form flux units through the B-cycle (i.e.,
the number of cascade steps).
Since each cascade step reduces the mass-scale of the theory by
a factor $e^{2\pi/(3 g_s M)}$ \refs{\Giddings,\HKO}, the string calculation
gives $\epsilon \sim M_{string}^3 e^{-2\pi l/(g_s M)}$,
in perfect agreement with \epsir.

\newsec{Turning on a Fayet-Iliopoulos Term}

In the previous sections we occasionally discussed
the effects of gauging $U(1)_{baryon}$ and of turning on a
Fayet-Iliopoulos parameter $\xi$.  Since we will use such a term
below, in this section we summarize and comment on these results.

Consider first the effect of gauging this symmetry with $\xi=0$.
The moduli space of vacua is the same as \quanm
 \eqn\quanma{\oplus_{r=1}^M\oplus _{l=0}^k
 Sym_{p-lM}(\CC_{r,l})}
except that for $p=kM$ the factor $Sym_0(\CC_{r,l=k})$ is a point
rather than a copy of $\IC$.  Also, the number of $U(1)$ factors
in the low energy theory is always given by $p-lM$; i.e.\ all the
moduli are in $\CN=4$ multiplets.

Now, let us consider a non-zero value for $\xi$.
For a small $U(1)_{baryon}$ gauge
coupling $g$, the effect of $\xi$ can be analyzed in the low energy
theory.  For generic values of $M$ and $p$, the low energy theory
is $U(1)^{p-lM}$ and there are no light charged fields.  $\xi$ is
the Fayet-Iliopoulos term of a particular linear combination of
these $U(1)$ factors.  Since there are no massless charged fields,
it is clear that supersymmetry is broken, and for small $g$ the
vacuum energy is
 \eqn\vacenxi{V={ g^2 \xi^2 \over 2}\ .}
This agrees with our classical answer \clasxi\ but this derivation
is more general because it includes also all the quantum
corrections due to the strong $SU(N_1)\times SU(N_2)$ dynamics.

There is only one exception to this result.  For $p=kM$ the theory
with $\xi=0$ has $M$ isolated vacua with $l=k$ containing no low energy
gauge fields.  In these vacua the $U(1)_{baryon}$ gauge symmetry
is Higgsed.  Therefore, turning on $\xi$ in these vacua does not
break supersymmetry, but instead it moves the vacuum in field
space.  It is important that even in this case of $p=kM$ the
result \vacenxi\ still applies to the other branches of the moduli
space with $l=0,...,k-1$.

Let us consider the specific example of the theory with $p=kM+1$
on the branch with $l=k$.  Different values of $k$ are related by
the duality transformations (the example of cascading from $k=1$
to $k=0$ was discussed in section 7).  Therefore, we can focus on
the simplest case, $p=1$, where we find an $SU(M+1) \times
U(1)_{baryon}$ gauge theory with the $A$ and $B$ fields. The moduli space is
described by a single D3-brane moving on the deformed conifold
$\CC_{r,0}$: in the gauge theory its position is encoded in the
meson fields ${\cal M}_{\alpha \dot\alpha}$.  We will consider the
leading order effect in the $U(1)_{baryon}$ gauge coupling $g$ far
out along the flat direction. There we can use the classical
approximation to find the potential
 \eqn\Dtermpot{ {1\over 2} g^2\left ( |{A}_{\alpha a}|^2 -
 |{B}_{\dot \alpha}^a|^2 -\xi\right )^2 \ .}
So, the potential of a moving D3-brane picks up a positive
constant shift $g^2\xi^2/2$.

We can also see this effect without gauging $U(1)_{baryon}$.
Consider the $p=M+1$ theory and separate a single D3-brane. As we
discussed at the end of section 4, here $SU(2M+1)\times SU(M+1)$
is broken to $SU(2M)\times SU(M)\times U(1)$.  So, in addition to
containing the meson $\CM_{\alpha\alphadot}$, our low-energy theory is the
$p=M$ theory with gauged $U(1)_{baryon}$, whose gauge coupling
originates from the non-Abelian gauge coupling $g_{YM}$. This
gauging removes the baryonic branch of this theory. But we could
attempt to move the fields in the direction of the baryonic branch
by turning on a nonzero value $\langle{\cal U}\rangle \sim U$ in
\ztwoop. Since the baryonic branch is lifted, this breaks
supersymmetry and leads to a potential for $\CM$ or order $U^2$.
In section 15, we will find the supergravity dual of this effect.
We will treat this mobile D3-brane as a probe and will calculate
its potential. The nonzero value for $U$ will appear because we will
place the D3-brane on a resolved warped deformed conifold.

\newsec{Matching Gauge Theory and String Theory }

In this section we would like to analyze various objects in the theory,
and compare their gauge theory and string theory
descriptions. These objects include domain walls, and confining and solitonic
strings.

\subsec{Domain Walls in the Gauge Theory}

In field theories, domain walls interpolate between different
vacua. What are the possible domain walls in the confining
$SU(M+p)\times SU(p)$ gauge theory? First, it is clear that there
are no domain walls interpolating between two different vacua on
the same branch.  Such a domain wall simply spreads out and
becomes infinitely thick.

Second, we examine domain walls interpolating between vacua on
different branches.  The most interesting case is when the wall
interpolates between two branches with different $r$ but with
the same value of $l$.  Branches with the same $l$ have a natural
one-to-one map between them.  Therefore, we consider a domain wall
which interpolates between a point in branch $r$ and its image in
branch $r'$.  As far as the low energy $SU(M)$ theory, this is a
familiar situation of a domain wall which interpolates between two of the
$M$ vacua produced by the breaking of the $\IZ_{2M}$ R-symmetry
to $\IZ_2$ \refs{\DvaliXE\Kovner-\WittenEP}.  Therefore, we learn that this
domain wall is BPS and its tension is
 \eqn\domtenbps{M\left|\Lambda(M,p,l)^3(e^{2\pi i r \over M}-
 e^{2\pi i r' \over M})\right|}
For large $M$ this becomes
 \eqn\domtenbpsl{M\left|\Lambda(M,p,l)^3(e^{2\pi i r \over M}-
 e^{2\pi i r' \over M})\right| \to 2\pi \left|\Lambda(M,p,l)^3
  (r-r')\right|}
Standard large $M$ counting has $\Lambda(M,p,l)^3 \sim M$
\WittenEP, and the tension of the domain wall is of order $M$.
Therefore, in the `t Hooft limit, this scales as a D-brane tension
\WittenEP. Indeed, in the string theory dual of our gauge theory
these domain walls are the D5-branes wrapping the $\IS^3$ at the
bottom of the deformed conifold $r-r'$ times
\refs{\KlebanovHB,\MN,\Acharya}.

The domain wall tension \domtenbps\ is independent of the moduli.
This is a general property of BPS domain walls.  Consider a BPS
domain wall which interpolates between a vacuum $a$ and a vacuum
$b$.  Its tension is $|W(a)-W(b)|$, where $W(a)$ and $W(b)$ are
the values of the superpotentials in the two vacua.  Now assume
that either the vacuum $a$, or the vacuum $b$ or both are on a
moduli space of supersymmetric vacua.  Then, it is clear that
$W(a)$ and $W(b)$ are independent of the moduli; otherwise, that
superpotential would have led to a potential along the moduli
space.  Since $W(a)$ and $W(b)$ are independent of the moduli, so
is their difference, which is the tension.  This simple argument
shows that the tension of a BPS domain wall
is independent of the moduli.

A slightly more complicated example is obtained from the one we
have just discussed by letting the domain wall interpolate between
two vacua which are not isomorphic.  It is clear that the lowest
energy configuration is obtained by first interpolating between
two isomorphic points, as above, and then interpolating to the
desired vacuum.  It is also clear that this second step in the
interpolation will make the wall spread out and make it non-BPS.

The most complicated example occurs when we try to interpolate
between vacua with different values of $l$.  Since there is no
one-to-one correspondence between branches of the moduli space with
different $l$, it is clear, by the argument above that such domain
walls cannot be BPS.  The most we can say about them is that their
tension is bounded by the difference in the superpotential
 \eqn\nsfb{\eqalign{
 T> &M\left|\Lambda(M,p,l)^3e^{2\pi i r \over M} -
 \Lambda(M,p,l')^3e^{2\pi i r' \over M}\right| \cr
 =&M\left|
 \Lambda(M,p,l=0)^3\left(I(M,p)^{l\over M}e^{2\pi i r \over M} -
 I(M,p)^{l'\over M} e^{2\pi i r' \over M}\right)\right|\cr
 \to & M\left| \Lambda(M,p,l=0)^3\left(e^{2\pi i \tilde \tau l} -
 e^{2\pi i \tilde \tau l'} \right)\right|
 }}
where we have used $I=e^{2\pi i \tau}= e^{2\pi i \tilde \tau M}$
and the fact that in the `t Hooft limit, $ \tilde \tau $ is of
order one. In the supergravity approximation, $|\tilde \tau| \ll 1$.
Then, we may expand
 \eqn\nsfba{
 T> M\left| \Lambda(M,p,l=0)^3\left(e^{2\pi i \tilde \tau l} -
 e^{2\pi i \tilde \tau l'} \right)\right|
 \approx 2\pi M\left| \Lambda(M,p,l=0)^3\tilde \tau (l-l') \right|
 \ .}
So, this tension is bounded from below by
order $M^2$. In the `t Hooft limit, this scales as an NS5-brane
tension. Indeed, these
domain walls are dual to the NS5-branes wrapping the $\IS^3$ at the bottom of
the deformed conifold \Kachru.

\subsec{Domain Walls in the Dual String Theory}

In the supergravity duals, the BPS domain wall separating
the adjacent vacua is a D5-brane
wrapped over the round 3-sphere at $t=0$ \refs{\KlebanovHB,\MN,\Acharya}.
In section 14.1 we will show that the tension of this wrapped
D5-brane does not depend on the baryonic branch modulus, in
agreement with the field theory considerations.
Therefore, to calculate the tension of the wrapped D5-brane, we will
work at the $\IZ_2$ symmetric locus on the baryonic branch,
$|{\cal A}|= |{\cal B}|$,
described by the warped deformed conifold solution \KlebanovHB.
Recall that the metric is
\eqn\metric{ ds_{10}^2 = H_{KS}^{-1/2} (t) dx^2 + H_{KS}^{1/2}
(t) ds_6^2\ , }
 where $ds_6^2$ is the Calabi-Yau metric on the deformed conifold
 \eqn\dc{ \sum_{i=1}^4 z_i^2 = \varepsilon^2 \ . }
 Its explicit form is
given, for example, in \KlebanovHB. At $t=0$ one finds a round
3-sphere of radius-squared $\varepsilon^{4/3} (2/3)^{1/3}$. Hence,
its volume is $2\pi^2 \varepsilon^2 \sqrt{2/3}$. The tension of
the domain wall is \eqn\sugratension{ T = \varepsilon^2
{\sqrt{2/3}\over 16 \pi^3 g_s (\alpha')^3} \ . } Note that powers
of $H_{KS}(0)$ cancel in this calculation, since the D5-brane has
three directions within $\IR^{3,1}$ and three within the deformed
conifold.

To match the string and field theory parameters,
we set \sugratension\ equal to the field theory result,
 \eqn\compare{ \Lambda(M,p,l)^3 \sim M {\varepsilon^2
 \over g_s M (\alpha')^3} \ .}
Since  both $\varepsilon$ and $g_s M$ are held fixed in the `t
Hooft limit, we see that $\Lambda(M,p,l)^3$ is of order $M$
\WittenEP.\foot{ Comparing with the conventions of \epsilonl,
where $\epsilon \sim \Lambda(M,p,l)^3$, we find $\epsilon\sim M
{\varepsilon^2\over g_s M(\alpha')^3}$. The fact that in the 't
Hooft limit $\epsilon \sim \Lambda^3$ scales as $M$, while
$\varepsilon $ is of order one can be traced back to the scaling
of the coordinates $\CM$ by $\sqrt h$ in \cmdeff.  Since in the 't
Hooft limit $h \sim M$, the deformation in terms of $\CM$ is of
order one.} Thus, the IR scale kept fixed in the large $M$ limit
is \eqn\newscale{ \tilde\Lambda(M,p,l) = M^{-1/3} \Lambda(M,p,l) \
,} and we find \eqn\sugramatch{ {\varepsilon^2\over (\alpha')^3}
 \sim g_s M  \tilde \Lambda(M,p,l)^3
\ .}

The cascading theory has another type of domain wall which
separates vacua with adjacent values of $l$. The dimensions of the
moduli space on the two sides of this domain wall are different,
hence this domain wall cannot be BPS saturated. Therefore, its
tension is not given by the difference between the values of the
superpotential. In the supergravity dual this domain wall is an
NS5-brane wrapped over the 3-sphere \Kachru. To see that this
identification is correct, we note that the $M$ units of RR flux
through the 3-sphere require that $M$ D3-branes end on the
NS5-brane (this is the Hanany-Witten effect \Hanany). Hence, upon
crossing the domain wall, we find $M$ additional D3-branes
corresponding to $l\rightarrow l-1$. This is why the dimensions of
the moduli space differ on the two sides of the domain wall. The
presence of the $M$ D3-branes attached to the wrapped NS5 makes it
difficult to define the domain wall tension: it is a boundary term
that must be separated from a much bigger bulk term related to the
back-reaction of the $M$ D3-branes filling $\IR^{3,1}$ on the
supergravity background. Hence, the calculation of the non-BPS
domain wall tension is a difficult task.

\subsec{Interpolation in $g_s M$ and other comparisons}

The large $M$ cascading gauge theory has a continuous parameter
$g_s M$.
For small $g_s M$ the spacing between cascade steps
is large \KlebanovHB,
corresponding to the `choppy' RG flow discussed in detail
in \StrasslerQS. Far in the IR such a theory is well-approximated
by the usual ${\cal N}=1$ supersymmetric
gluodynamics;
hence, we expect the square root of the
confining string tension and the glueball masses to be of order
$\left|\tilde \Lambda(M,p,l)\right|$.
Let us compare these results with the supergravity predictions,
which are valid for large $g_s M$ \HKO:
\eqn\stringsugra{
T_s^{1/2} \sim {\varepsilon^{2/3}\over \alpha'\sqrt{g_s M} }
\sim
{\left|\tilde \Lambda(M,p,l)\right|\over (g_s M)^{1/6} }\ ,
}
\eqn\mglue{ m_{glueball}\sim
{\varepsilon^{2/3}\over \alpha'g_s M }
\sim
{\left|\tilde \Lambda(M,p,l)\right|\over (g_s M)^{2/3} }\ .
}
More generally, we have
\eqn\moregen{
T_s^{1/2} \sim
\left|\tilde \Lambda(M,p,l)\right| f_s(g_s M)\ ,\qquad
m_{glueball}\sim
\left|\tilde \Lambda(M,p,l)\right| f_g(g_s M)\ ,}
where $f_s (g_s M)$ interpolates between a value of order one
at small $g_s M$,
and the $(g_s M)^{-1/6}$ fall-off at large values.
Similarly, $f_g$ interpolates between values of order one and
the $(g_s M)^{-2/3}$ fall-off. Interpolations of this sort are typical
in gauge/gravity dualities.

Now, let us discuss D-branes at the bottom of the warped deformed
conifold. In the probe approximation, the D-string tension is
\eqn\dstring{ {1\over 2\pi \alpha' g_s H_{KS}(0)^{1/2}} \sim M
{\left|\tilde \Lambda(M,p,l)\right|^2\over (g_s M)^{4/3} }\ . }
Note that it is proportional to $M$. In the non-compact conifold
case, there is a problem with the probe approximation. The
D-string introduces a monodromy of the massless pseudoscalar mode
\GubserQJ, which causes an IR logarithmic divergence. In the SUGRA
calculation this divergence comes from the perturbation $\delta
F_{01r}$ introduced by the string stretched in the $x^1$
direction. Thus, to discuss the tension we have to introduce an IR
cut-off. However, if we embed the deformed conifold in a string
compactification, then the $U(1)_{baryon}$ symmetry is gauged, and
the IR divergence is removed.

The D-string at the bottom of the warped deformed conifold should
be dual to a solitonic string in the cascading gauge theory, which
couples to the Goldstone boson \GubserQJ.
The field theory discussion of such solitonic strings again
presumes either an IR regulator, which removes the logarithmic
divergence in the tension, or a gauging of
$U(1)_{baryon}$ which turns the string into a string of
Abrikosov-Nielsen-Olesen type.
On general grounds, the tension of this string
should satisfy
\eqn\gensolit{ T_{soliton}= M
\left|\tilde \Lambda(M,p,l)\right|^2 f_{soliton}^2(g_s M) \ , }
where $f_{soliton}(g_s M)$ falls off as $(g_s M)^{-2/3}$ at
infinity. Note that there is no such soliton in the ${\cal N}=1$
supersymmetric $SU(M)$ gauge theory. Therefore, we expect $f_{soliton}$ to
diverge as $g_s M\rightarrow 0$.

Another very interesting non-BPS object is the anti-D3 brane,
whose tension is
\eqn\antidthree{ {1\over 8\pi^3 (\alpha')^2 g_s H_{KS}(0)} \sim
M {\left|\tilde \Lambda(M,p,l)\right|^4\over (g_s M)^{5/3} }\ .
}
At a general coupling, we expect the energy of this excitation per unit volume
 to behave as
\eqn\gensolit{
M \left|\tilde \Lambda(M,p,l)\right|^4 f_D^4(g_s M)
\ .
}
 Again, this object might not be present in the pure supersymmetric
gluodynamics; therefore, $f_D(g_s M)$ should blow up near zero.
Thus, this is a very heavy object in the limit of widely spaced cascade steps.
The smallest theory where it may exist is
$SU(2M) \times SU(M)$ that appears at the bottom of the cascade.
The fact that the tension scales as $M$ suggests that only one eigenvalue of
the meson matrix ${\cal M}$ is excited.

\newsec{Supergravity Dual of the
Cascading Theory on the Baryonic Branch}

In this and the subsequent sections we review the dual
supergravity description of the baryonic branch of the cascading
$SU((k+1)M)\times SU(kM)$ gauge theory, and compare various
supergravity observables with the gauge theory along this branch.
The simplest gauge theory picture of the baryonic branch is found
in the far infrared $SU(2M)\times SU(M)$ theory where
\eqn\simpbar{ {\cal A}= i\Lambda_1^{2M}\zeta\ ,\qquad {\cal B}=
i\Lambda_1^{2M}/\zeta\ ,} and $\zeta$ is the complex modulus for
the branch. The gauge theory with $|\zeta|=1$ is described by the
warped deformed conifold solution of \KlebanovHB. The gauge theory
has a pseudoscalar Goldstone mode \AharonyPP\ corresponding to
changes in the phase of $\zeta$. Its supergravity dual was
constructed in \GubserQJ. This mode vanishes at zero momentum, in
agreement with the fact that the Goldstone boson has only
derivative couplings. Therefore, position-independent changes of
the phase of $\zeta$ do not produce any new supergravity
backgrounds.

The baryonic branch of the supergravity backgrounds is labelled by
a real parameter $|\zeta|$. In \GubserQJ\ it was proposed that
this branch falls within the Papadopoulos-Tseytlin (PT) ansatz
\PT\ for backgrounds of IIB SUGRA describing the deformed conifold
with fluxes. The ten-dimensional  metric of the PT ansatz
is\foot{Following \Butti, we use this ansatz for the string frame
metric.} \eqn\Pap{\eqalign{& ds^2_{10} =  H^{-1/2}  dx_m dx_m +
e^x ds_6^2 ,\cr & ds_6^2 =  (e^{g}+ a^2 e^{-g})  ( e_1^2 + e_2^2)
      +   e^{-g}  \sum_{i=1}^2 \left ( \epsilon_i^2 - 2a e_i\epsilon_i \right )  +
v^{-1} (\tilde{\epsilon}_3^2+dt^2)\ ,
}}
where $H,x,g,a,v$ are functions of the radial variable $t$. The
definitions of the 1-forms, and the ansatz for $H_3,F_3,F_5$ are
reviewed in Appendix A; we ask the reader to refer to the notation
there. While the necessary backgrounds are quite complicated, they
simplify considerably in the large radius (UV) limit, where they
approach the asymptotic cascade form found in \KT. This asymptotic
may be approximated by $AdS_5\times T^{1,1}$ modulo slowly-varying
logarithms \KT\ which are present due to the logarithmic RG flow
in the dual gauge theory \KN.

The PT ansatz is $SU(2)\times SU(2)$ invariant but in general
breaks the $\IZ_2$ symmetry that interchanges the two $\Bbb S^2$'s
of $T^{1,1}$. In the field theory the corresponding symmetry is
the interchange of $A_\alpha$ with $B_{\dot\alpha}$ accompanied by charge conjugation
in both $SU(N_1)$ and $SU(N_2)$ \KW. This $\IZ_2$ symmetry is
restored for the warped deformed conifold solution of \KlebanovHB\
corresponding to $|\zeta|=1$. Since the breaking of this discrete
symmetry is associated with the resolution of the conifold, the
solutions with broken $\IZ_2$ may be called {\it resolved warped
deformed conifolds}.

The PT ansatz was originally introduced in search of an
extrapolation between the warped deformed conifold (KS) background
\KlebanovHB, which preserves the $\IZ_2$ symmetry, and the
Maldacena-Nunez (MN) background \MN\ which breaks it. In
\GubserQJ\ the linearized deformations around the KS background,
which are $\IZ_2$ odd, were found using the PT ansatz. They were
interpreted as the supergravity duals of small motions along the
baryonic branch of the cascading gauge theory, corresponding to
$|\zeta| \approx 1$. It has been conjectured that far along the
baryonic branch the background approaches the MN background \IK.
However, this cannot be true far in the UV since the MN background
asymptotes to a linear dilaton rather than to the KT solution \KT.
Subsequently, Butti, Gra\~ na, Minasian, Petrini and Zaffaroni
(BGMPZ) wrote a remarkable paper \Butti, where the method of
$SU(3)$ structures was used to derive a system of coupled
first-order equations for the functions $a(t)$ and $v(t)$,
describing an ${\cal N}=1$ supersymmetric solution to the PT
ansatz. The solution of these equations determines other unknown
functions (see Appendix B), so that the problem of constructing
the family of supergravity duals of the entire baryonic branch
became tractable, at least numerically. It turns out that the
backgrounds far along the baryonic branch do approach the
appropriately shifted MN solution {\it in the IR}, yet in the UV
they have the cascade asymptotics of \KT.

\subsec{Relation between the warp factor and the dilaton}

We will be particularly interested in the the dilaton profile
$\phi(t)$ and the warp factor $H(t)$ which determine the tensions
of many probe branes. In our conventions, the position-dependent
string coupling is $g_s e^{\phi(t)}$, and
we set $\phi(\infty)=0$. The dilaton profile is
determined by \Butti
\eqn\phieq{ \phi' =  {\left( C-b\right) {\left(
a\,C-1 \right) }^2 \over \left( b\,C-1 \right) S}\,e^{-2\,g}\ , }
with the definitions of functions $b,C,S$ given in Appendix B.
 The equation for the warp
factor may be written in the form
\eqn\warpfact{
H' =  -K(t) e^{-2x(t)} H(t) \ , }
which implies
that the self-dual 5-form field strength is
\eqn\selfd{g_s F_5 =
d\left (H^{-1}\right )\wedge d^4x + K(t) e_1 \we e_2 \we
\epsilon_1 \we \epsilon_2 \we \epsilon_3\ , } i.e.
$g_s C_{0123}=
H^{-1}(t)$.
Using \phieq\ and formulae in Appendix B, we find that
$$K(t) e^{-2x(t)}={2\phi'\over 1- e^{2\phi(t)}}
\ .
$$
Hence, \warpfact\ may be written in the form
\eqn\warpfactnew{
H' =  -{2\phi'\over 1- e^{2\phi(t)}} H(t) \ . }
This may be integrated to give
\eqn\A{ H(t)=\tilde H\left (e^{-2\phi(t)} -1\right ) \
\ ,}
where $\tilde H$ is an integration constant.
To achieve a decoupled field theory in
gauge/gravity dualities, one requires that the warp factor
$H(t)$ vanishes at infinity. Since $\phi(\infty)=0$,
\A\ clearly satisfies this requirement for any
$\tilde H$.
A more detailed analysis of the boundary conditions at large $t$,
which will allow us to determine $\tilde H$,
will be presented in the next section.

\newsec{Boundary Conditions and Analysis of Solutions}

To specify the solution completely, we need to fix the boundary
conditions in the UV region $t\to \infty$. In order to find the
correct boundary conditions on the solutions along the baryonic
branch, let us recall the $\IZ_2$ symmetric KS solution
\KlebanovHB.\ In terms of the PT variables, this solution has
\eqn\aks{\eqalign{ & a_{KS}=-{1\over \cosh(t)}\ ,\cr &
v_{KS}={3\over 2}\left(\coth (t) - {t\over \sinh^2(t)}\right)\
,\cr & e^{g_{KS}}=\tanh t \ ,\cr & \phi_{KS}=0\ ,\cr & e^{-4
A_{KS}(t)}=H_{KS}(t)=2^{-8/3} \gamma I(t)\ ,}} where we define
\eqn\warpdefin{I(t)= \int_t^\infty dx{x\coth x-1\over \sinh^2 x
}(\sinh 2x-2x)^{1/3}\ , \qquad \gamma= 2^{10/3} (g_s M\alpha')^2
\varepsilon^{-8/3}\ .} One finds \HKO\ that $I(0)\approx 0.71805$,
while for large $t$,
$$I(t)\to 3\cdot 2^{-7/3} (4t-1) e^{-4t/3}+ \ldots\
$$
The large $t$ expansion of the
warp factor is therefore given by
\eqn\huv{ \gamma^{-1} H(t)= {3\over 32}e^{-4t/3}(4t-1)- {25
t^2 - 85 t+ 12\over 125} e^{-10 t/3} + O\left (e^{-16t/3} \right )
\ .}

Moving along the baryonic branch away from the $\IZ_2$ symmetric
solution of \KlebanovHB\ corresponds to changing expectation
values of fields in the cascading gauge theory. In the dual
supergravity description, such changes typically preserve the
leading asymptotics of the fields but affect the sub-leading terms
\KWnew. This is the standard fact for asymptotically AdS spaces,
and is expected to apply also to the cascading case where the UV
asymptotics differ from AdS only by logarithmic corrections.
Thus, for the entire baryonic branch of solutions we will require
that the leading asymptotics are the same as in the KS case, i.e.
$a(t) \to -2 e^{-t}$, $\gamma^{-1} H(t) \to {3\over
32}e^{-4t/3}(4t-1)$, etc. Similarly, we require that
$\phi(\infty)=0$.

\subsec{Expansion around the KS solution}

The baryonic branch solutions that break the $\IZ_2$ symmetry
slightly were found in \GubserQJ: \eqn\ll{ a(t)= a_{KS}(t) (1 -
2^{-5/3} U Z(t)) + O(U^2)\ , \qquad e^g = e^{g_{KS}} (1- 2^{-5/3}
U Z(t)) + O(U^2)\ , } where \eqn\ll{ Z(t) = {\tanh t - t\over
(\cosh t\sinh t -t)^{1/3} } \ . } Thus, the asymptotic expansion
of $a(t)$ is\foot{ The parameter $U$ coincides with $a_{UV}$
introduced in \Butti. We use $U$ here rather than $a_{UV}$ to
stress the fact that $U$ is proportional to the expectation value
of the operator ${\cal U}$ \ztwoop. Therefore, $U$ parameterizes
the IR physics rather than UV: it is the modulus of the vacuum on
the baryonic branch.} \eqn\auv{ a(t)=-2e^{-t}+ U e^{-5t/3}(-t+1)+
\ldots} and $U$ parameterizes the resolution of the conifold
\KWnew, which breaks the $\IZ_2$ symmetry. In the gauge theory
this parameter is proportional to the expectation value of the
$\IZ_2$ odd operator ${\cal U}$ \ztwoop. The corresponding metric
component measures the difference between the radii-squared of the
$(e_1,e_2)$ two-sphere and the $(\epsilon_1,\epsilon_2)$
two-sphere: 
\eqn\ll{ e^x \left (e^g + e^{-g} (a^2-1)\right ) \sim U (t\coth t
-1) H_{KS}^{1/2} (t) + O(U^2 ) \ .} 
For large $t$, this falls off
as $t^{3/2} e^{-2t/3}\sim 
\varepsilon^{4/3} (\ln r)^{3/2} r^{-2}$ (here $r\sim
\varepsilon^{2/3} e^{t/3}$ 
is the radial variable of the near-AdS asymptotic \KT).
This is in agreement with ${\cal U}$ having dimension 2 \KWnew.
Hence, the expectation value of the operator may be read off from
the coefficient of the leading asymptotic 
(see \AharonyZR\ for a study of one-point
functions in the cascading background): 
\eqn\caluvev{\langle {\cal U}
\rangle\sim M U {\varepsilon^{4/3}\over (\alpha')^2}\ .} 
The expectation value of
${\cal U}$ is related to $\zeta$ (see \simpbar) through 
\eqn\ll{
\langle {\cal U} \rangle \sim M\Lambda_1^2 \ln |\zeta|  \ . }
Therefore, \eqn\urel{U\sim \ln |\zeta|\ . }

As shown in \Butti, the UV asymptotic expansions of $\phi$ and $H$
are
\eqn\phiuv{  \phi(t)=-{3\over 64} U^2
e^{-4t/3}(4t-1)+O\left (U^4 e^{-8t/3}\right ) \ ,}
\eqn\huvnew{ \gamma^{-1}H(t)=
{3\over 32}e^{-4t/3}(4t-1) -{3\over 32\cdot 512}
U^2 (256 t^3 - 864 t^2 + 1752 t-847) e^{-8t/3} + O\left
(e^{-10t/3}\right )\  }
 Comparing them with
\A, we find that
\eqn\Anought{ \tilde H= \gamma U^{-2} \ .}
In fact, taking $U$ to zero in \A, we find the expression
\eqn\phiall{ \phi(t)=-2^{-11/3} U^2 I(t) +O\left
(U^4\right ) }
valid for all $t$; i.e. the $O(U^2)$ term in $\phi(t)$
is proportional to $H_{KS}(t)$.
The UV expansion \phiuv\
is reproduced by \huv.

Moving along the baryonic branch corresponds to changing
$U\sim \ln |\zeta|$. It is also useful to parameterize the branch using the
parameter $y=y(U)$ which is defined via the IR expansion \Butti
\eqn\xidef{
a=-1+ \left ({1\over 2}+ {y\over 3}\right ) t^2+...
}
Comparing with the notation in the BGMPZ paper \Butti,
\eqn\ll{y= 3\xi_{BGMPZ}-3/2\ ,
}
but we will reserve $\xi$ for denoting the Fayet-Iliopoulos term.
Now, $y\in (-1,1)$ and the $\IZ_2$ simply acts as $y\to -y$.
Thus, at the KS point $U=y=0$,
corresponding to $|{\cal A}| = |{\cal B}|$. As $|{\cal A}| \to 0$,
we instead have $y\to -1$ and $U\to -\infty$; in this
limit the MN solution is approached in the IR, but the UV
boundary conditions correspond to the cascade rather than the
linear dilaton. $y$ may be determined as a function of
$U$ through numerical integration.

\subsec{Behavior far along the baryonic branch}

The initial idea motivating the PT ansatz was that it may
interpolate between the KS and the MN solutions. For the MN
solution corresponding to $y=- 1$ or $U\to -\infty$, 
\eqn\amn{ a_{MN}=-{t\over \sinh(t)}\ ,\qquad
v_{MN}=\sqrt{-1+2t\coth t-{t^2\over \sinh^2 t} }  \ . } We see
that this does not have the asymptotics \auv, which indicates that
$y=-1$ is a singular point which has to be excluded from the
baryonic branch. However, the solution can be arbitrarily close to
this point and still lie on the baryonic branch. In fact, far
along the baryonic branch the solutions become close to the MN
solution in the IR, but strongly depart from it in the UV: in the
UV all baryonic branch solutions have the ``cascading'' KT
asymptotics that are $AdS_5\times T^{1,1}$ modulo slowly varying
logarithms, while the MN solution asymptotes to the linearly
rising dilaton: \eqn\ll{ e^{2\phi_{MN}} \sim \sinh t \left (-1+ 2t
\coth t - {t^2\over \sinh^2 t}\right )^{-1/2} \ . }

For the solution on the baryonic branch, we instead fix
$\phi(\infty)=0$. Then the IR value of
the dilaton field, $\phi(t=0)=\phi_0$, starts from $0$ for $U=0$ and
approaches $-\infty$ for $|U|\to \infty$. Therefore, $\phi_0$
and $U$ are both zero in the KS case and approach minus
infinity in the MN limit $y\to -1$. We will later show that for large
$|U|$,
$e^{-\phi_0}\sim |U|^{3/4}$;
i.e. the effective string coupling is much weaker in the
IR than in the UV.\foot{If instead of keeping
$g_s$ fixed, we take a double scaling limit
where $U\to -\infty$, and $g_s\sim |U|^{3/4}$,
then we recover the MN solution \Butti.}
This means that for $|U|$ so large that $g_s M e^{\phi_0}$
becomes small, the supergravity background becomes highly curved in
the IR and
cannot be trusted. This follows from the fact that the
radius-squared of the $\IS^3$ at $t=0$ is of order
$\alpha' g_s M e^{\phi_0}$.

\fig{Plots of $a(t)$ and $v(t)$. The KS plot is shown in red,
$y=-3/4\ (U\approx -3.3)$ in green, $y=-0.99\ (U\approx -20.1)$ in blue, and MN
($y=-1$) in black.}{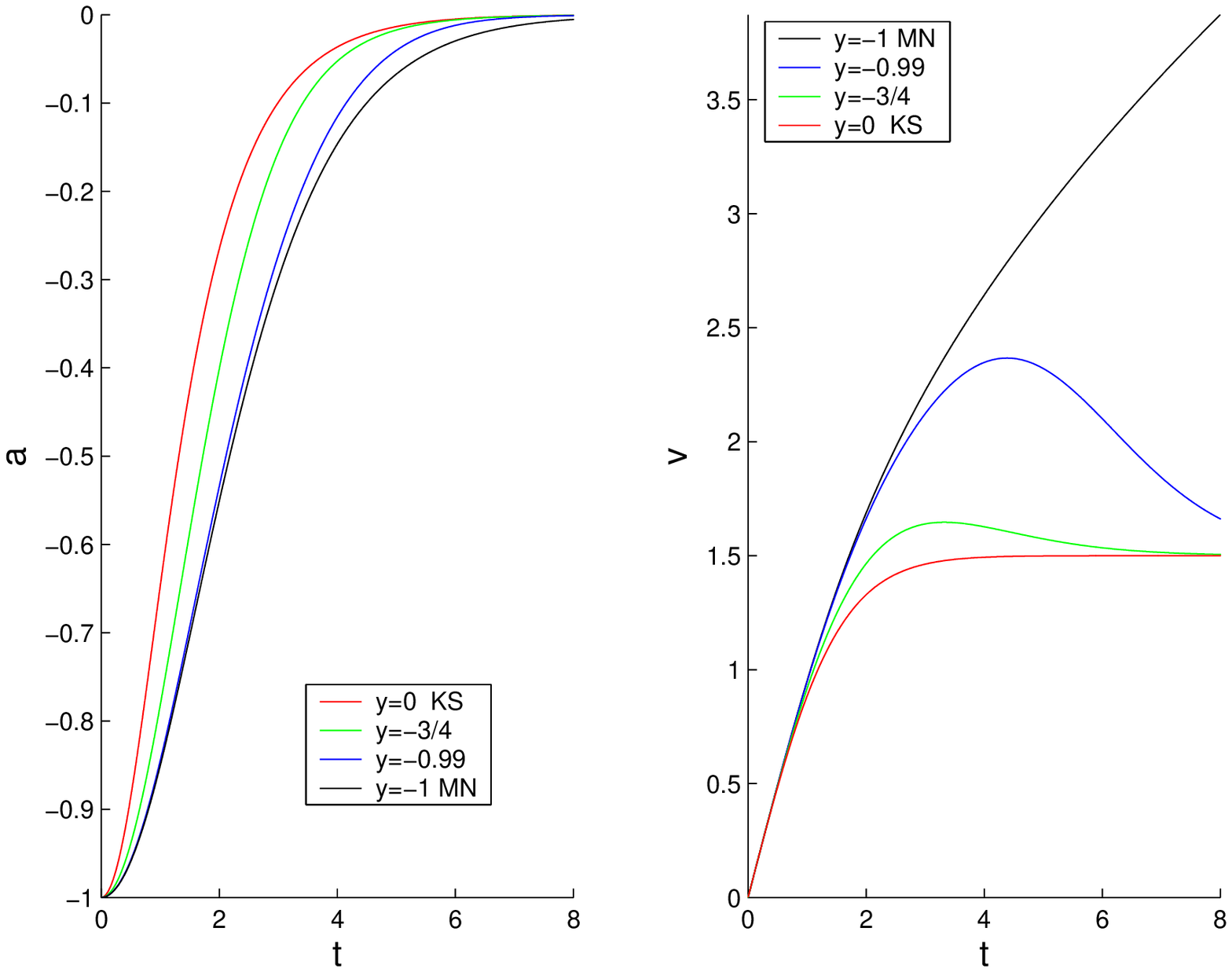}{9.5cm} \figlabel\seventeen

In Figures 1 and 2 we present the plots of $a(t),v(t)$ and
of $\phi(t),H(t)$ for the
KS ($y=0$), MN ($y=-1$) and intermediate values $y=-3/4\ (U\approx
-3.3)$ and $y=-0.99\ (U\approx -20.1)$.

\fig{Plots of
$\phi(t)$ and $H(t)$. The KS plot is shown in red,
$y=-3/4\ (U\approx -3.3)$ in green, $y=-0.99 \
(U\approx -20.1)$ in blue.}{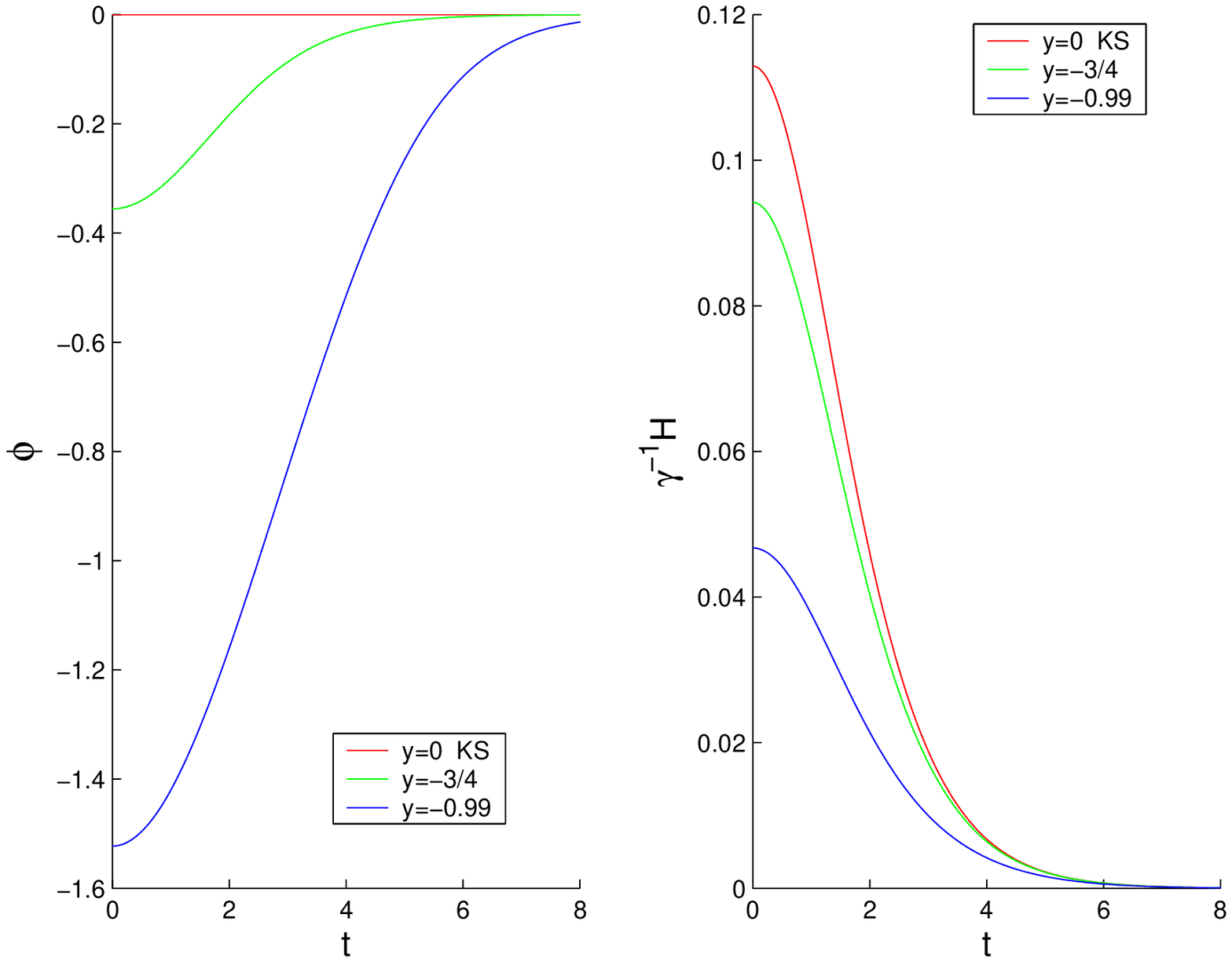}{9.5cm}

Figure 3 contains the profile of the dilaton for $y=-0.99$
and the exact MN dilaton shifted in a way that it starts from the
same value at $t=0$ as $\phi_{y=-0.99}$. The two graphs are
almost identical near $t=0$ but $\phi_{y=-0.99}$ quickly
approaches zero while
 $\phi_{MN}$ grows without bound.
The plots of $\phi(t)$, $a(t)$ and $v(t)$ show that even the $y=-0.99$ solution
approximates the MN solution well only for $t$ up to around $2$.
More generally, one can argue that, as $y\to -1$, the solutions approximate
the MN solution up to $t\sim -\ln (1+y)$. Thus, the
approach of the
IR behavior to that of the $MN$ solution as $y\to -1$ is logarithmically slow.

\fig{The red line is the dilaton profile for $y=-0.99\
(U\approx -20.1)$. The blue line is the $MN$ dilaton profile
shifted in a way that it starts from the same value at $t=0$ as
$\phi_{y=-0.99}$. }{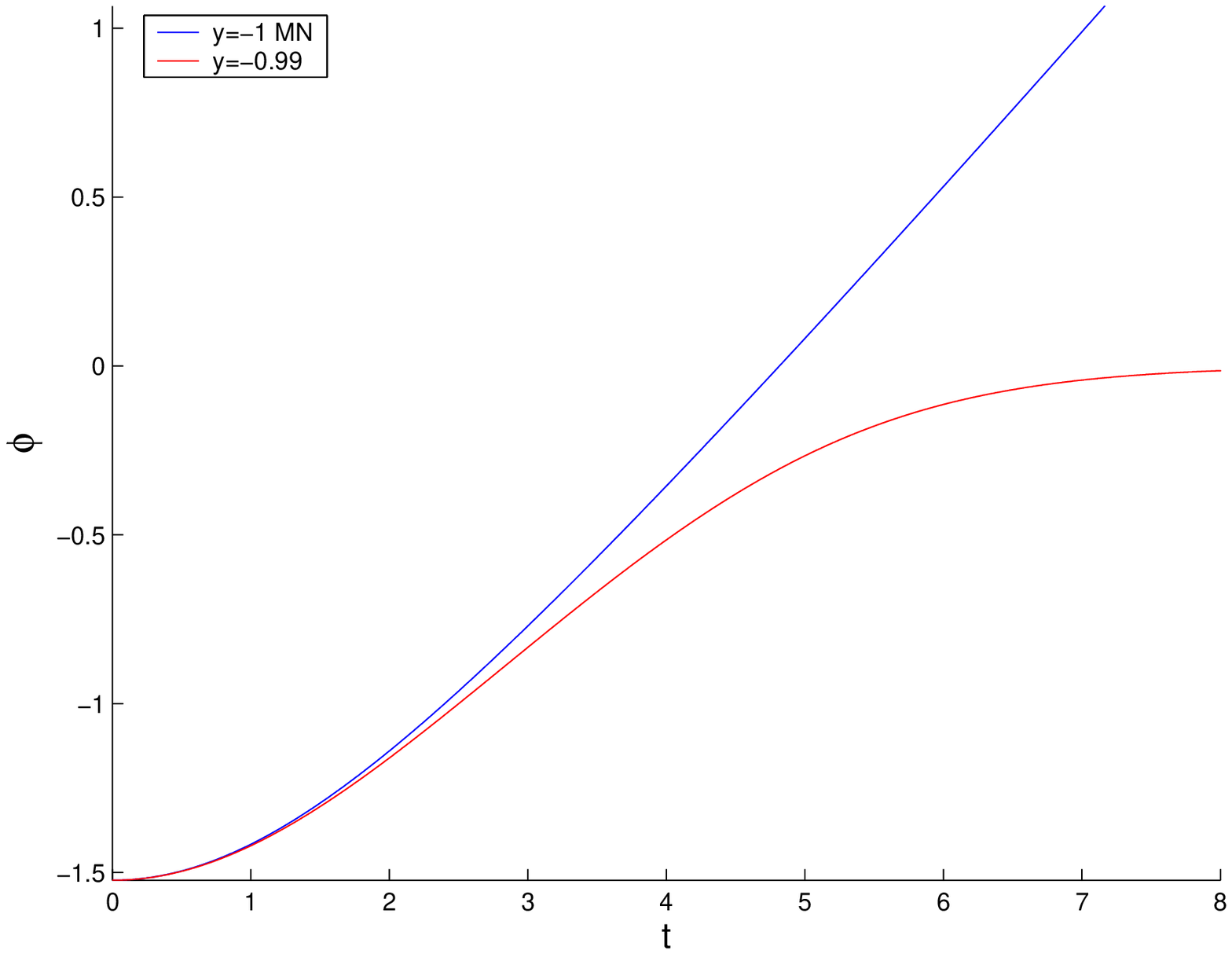}{7.5cm}

\newsec{The IR physics}

The IR physics is governed by the geometry near the origin
$t\rightarrow 0$:
\eqn\IRmetric{\eqalign{ &ds_{10}^2 =\ H_0^{-1/2} dx^2\ +\
{e^{\phi_0}\lambda\over 2} \left(dt^2+g_5^2+2g_3^2+2g_4^2\right) +
O(t^2)\ , \cr &g_5=\tilde{\epsilon}_3,~~~ \ \ \ \
g_3={e_1+\epsilon_3\over \sqrt{2}},~~~ \ \ \ \ \
g_4={e_2+\epsilon_4\over \sqrt{2}}\ ,\cr &H_0=\gamma
U^{-2}\left(e^{-2\phi_0}-1\right)
, \ \ \ \ \ \ \ \ \
\lambda^2=y^{-2}(1-e^{2\phi_0})\ . }}
We see that in
the far IR region the geometry is just $\mathbbR^{3,1} \times
\mathbbS^3\times \mathbbR^3$. The radius-squared of\  $\mathbbS^3$\
is $R^2=e^{\phi_0}\lambda$.

\subsec{Tension of the BPS Domain Wall}

\fig{The tension of a wrapped D$5$ brane}{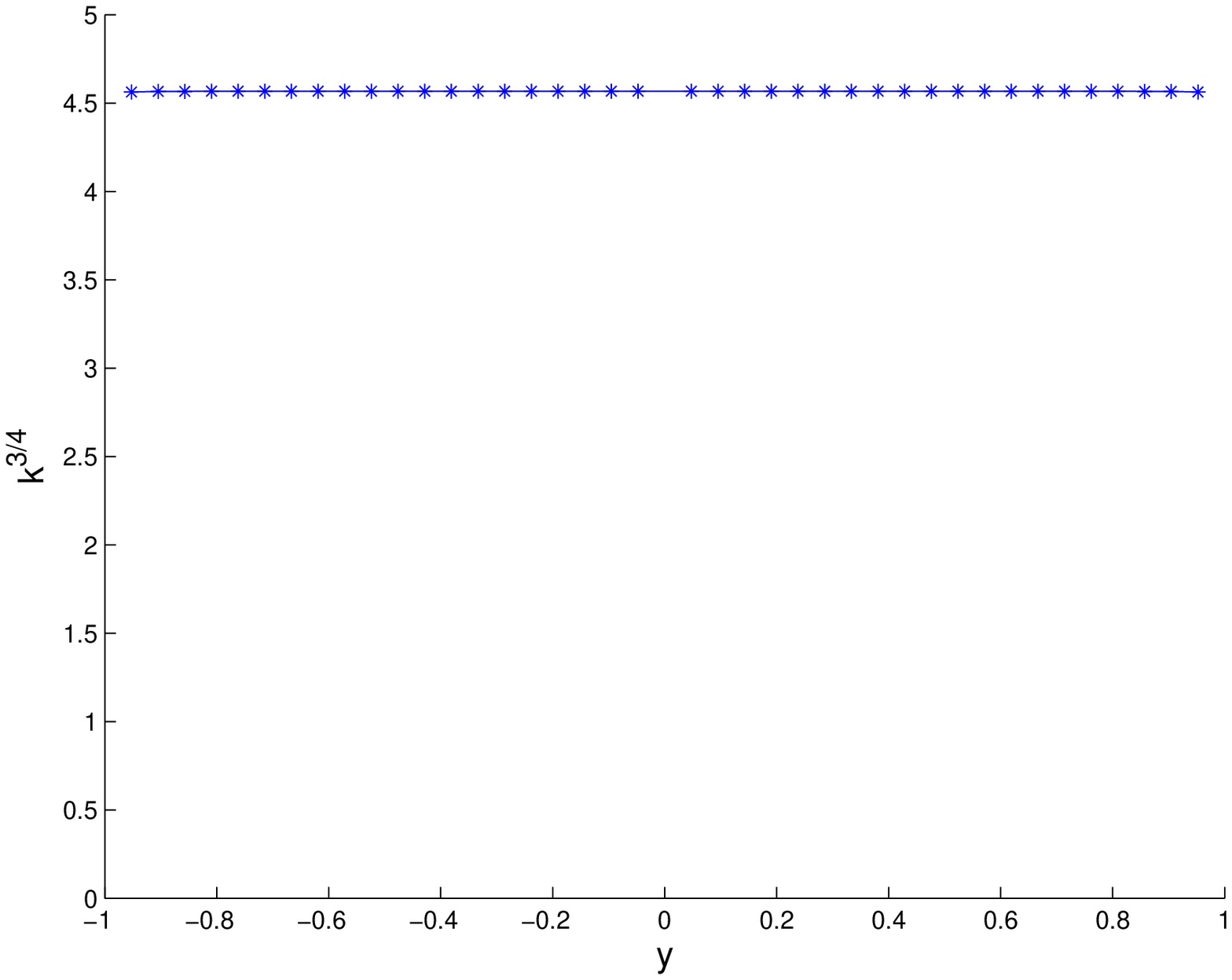}{9cm}\figlabel\dfive A D5-brane
wrapped over the round $\mathbbS^3$ at $t=0$ is the BPS domain
wall separating two adjacent vacua (there are $M$ inequivalent
vacua corresponding to the phase of the gluino condensate). This
is well-known to be a BPS object in the gauge theory, and we will
see a reflection of this in the dual string theory: the tension
does not depend on the baryonic branch parameter $U$.

The tension of the wrapped D$5$-brane is
\eqn\ll{ T={1\over 2
(2\pi\alpha')^3 g_s } H_0^{-3/4} e^{\phi_0/2}\lambda^{3/2}=
{1\over 2\  (2\pi\alpha')^3g_s} H_0^{-3/4}e^{\phi_0/2}
|y|^{-3/2}\left (1-e^{2\phi_0} \right )^{3/4}  \ .}
A numerical plot of this quantity as a function of $U$ is given
in Figure \dfive. It is constant within the numerical precision of the
calculation.
This is a nice check
of the boundary conditions we have imposed on the supergravity solution.

Let us
introduce $k$ via
\eqn\tdfive{ 2g_s (2\pi\alpha')^3 T=\gamma^{-3/4}
k^{3/4}={H_0^{-3/4}e^{\phi_0/2}\lambda^{3/2}}\ . } The value of
$k$ is easy to find at the KS point where, according to
\refs{\KlebanovHB,\HKO} \eqn\ll{ \lambda_{KS}=\left(k
H_{KS}(0)\right)^{1/2}= 6^{-1/3}2 I(0)^{1/2}=0.93266\ , } and
therefore \eqn\ll{ k=2^{4} 3^{-2/3}\ . } The irrational constant
$I(0)$ cancels because at the KS point all dependence on $H(0)$
cancels for such a wrapped brane: it has 3 directions within the
conifold and 3 directions within $\IR^{3,1}$. This is indicative
of the BPS nature of the wrapped D5-brane.

The constancy of $k$ provides us with a relation between $U$ and
the quantities $\phi_0$ and $y$ which are determined through
integrating the equations from large $t$ to $t=0$: \eqn\akxi{
U^2=k y^2 e^{-8/3 \phi_0}\ .} At large $|U|$, $|y|$ approaches $1$.
Hence, using \akxi, we see that $e^{-\phi_0}$ scales as
$|U|^{3/4}$. Using \akxi, we also find \eqn\ll{
H_0=\gamma{e^{-2\phi_0}-1\over U^2}={y^{-2}\gamma\over  k}
e^{2\phi_0/3}\left({1-e^{2\phi_0}}\right) \ .} This implies that
$H_0\sim |U|^{-1/2}$ for large $|U|$.

\subsec{Tensions of the fundamental string and anti-D3 brane}

The
dual of the confining string is the fundamental string placed at
$t=0$. As follows from \IRmetric, its tension is \eqn\ll{
T_s={1\over 2\pi \alpha'}H_0^{-1/2}\ . } This is not constant along
the branch, in agreement with the fact that
the confining string is not BPS saturated.
 Using \tdfive, we have
\eqn\ll{ H_0^{-1/2} = \gamma^{-1/2}k^{1/2} e^{-\phi_0/3}
\lambda^{-2} \ . } \fig{The confining string
tension}{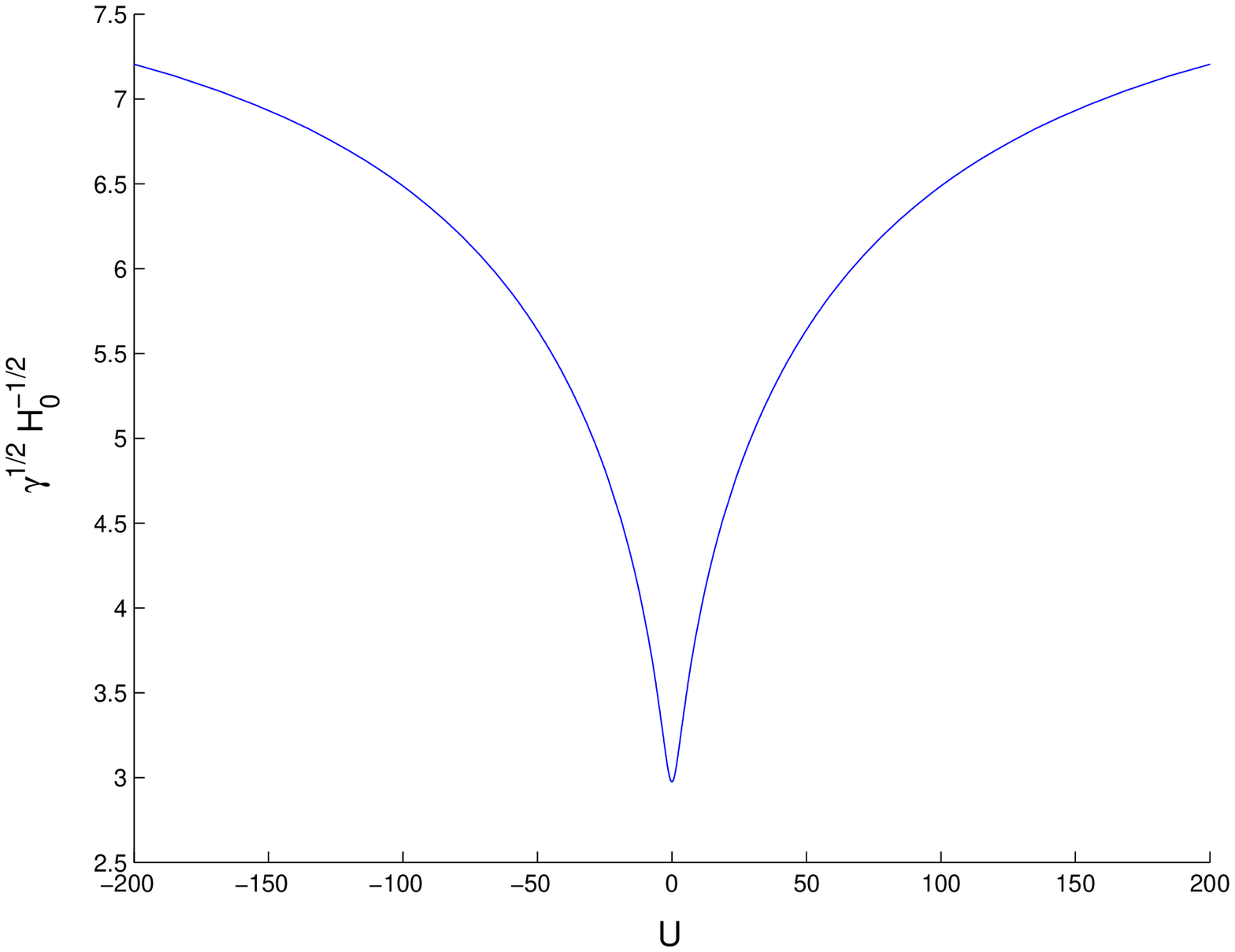}{9cm}\figlabel\ts At large $|U|$, $\lambda$
approaches $1$; hence, $T_s$ diverges as $e^{-\phi_0/3}\sim
|U|^{1/4}$. Figure \ts\ shows $T_s$ as a function of $U$. We have
also calculated some glueball masses along the baryonic branch,
and we find that they again diverge as a positive power of $|U|$.
We postpone a detailed presentation of the glueball results to a
future publication. \fig{The anti-D3 brane
tension}{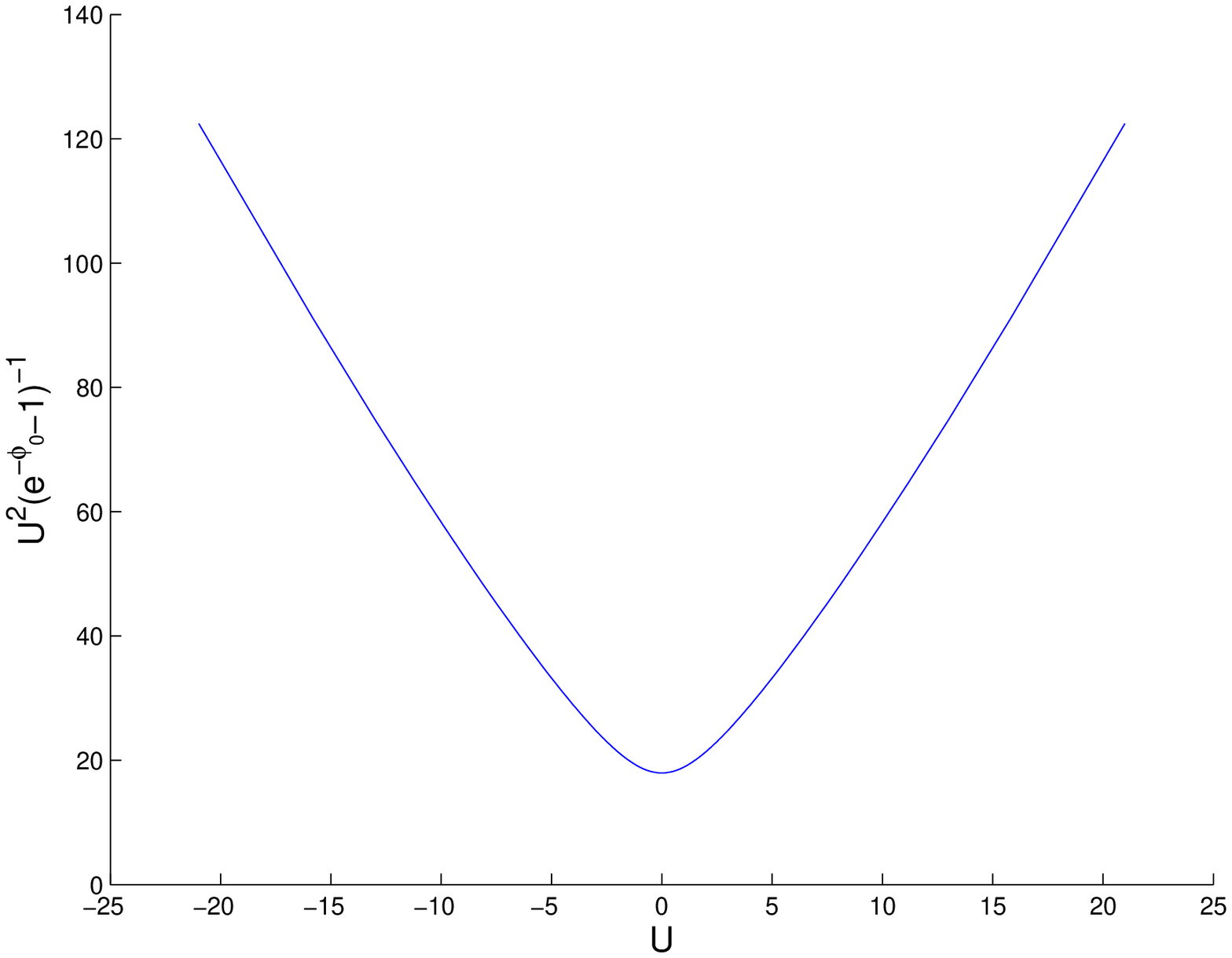}{9cm}\figlabel\antidtri \fig{The D3 brane
tension}{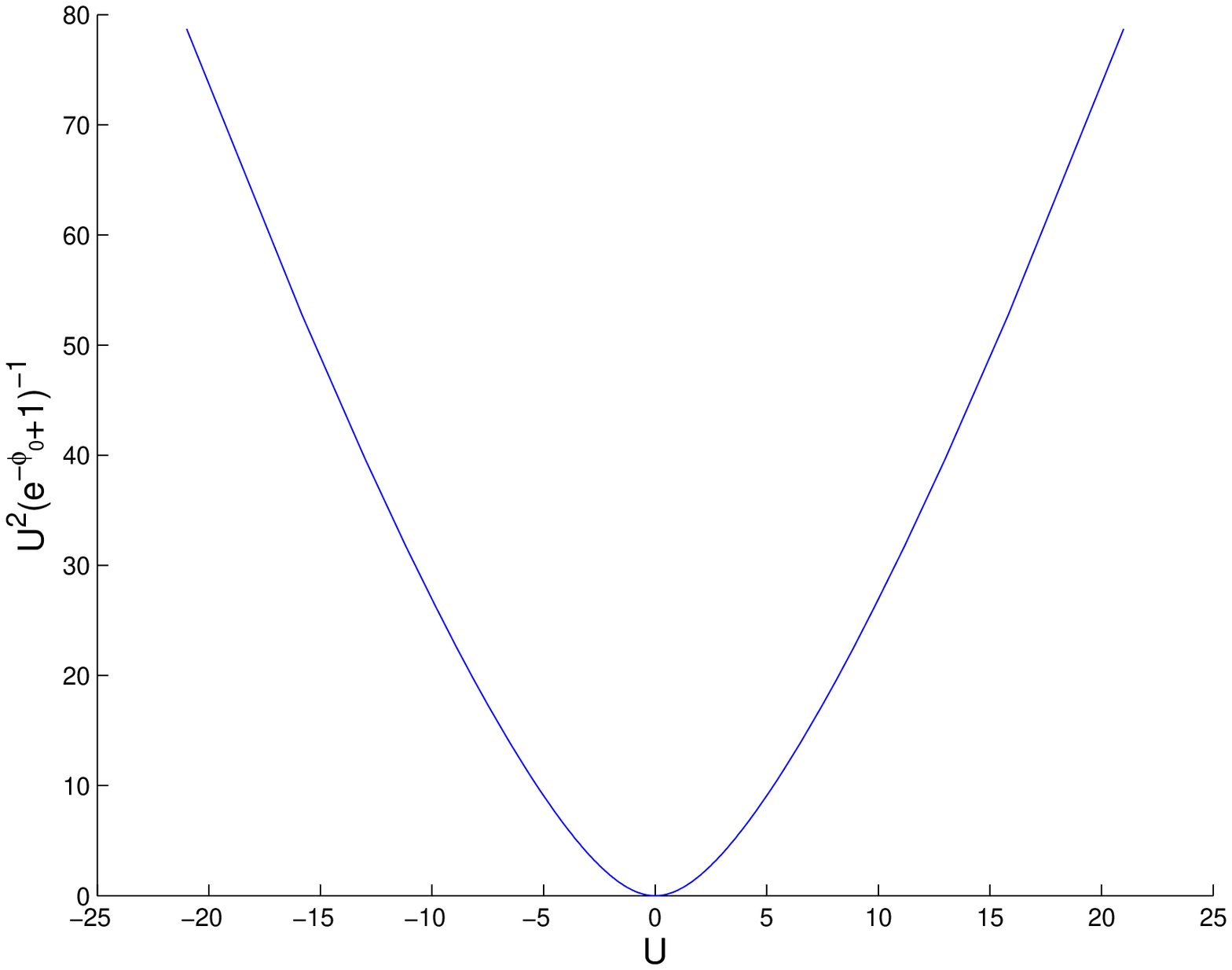}{9cm}\figlabel\dtri

Now consider an
anti-D3-brane parallel to $\IR^{3,1}$.
\foot{We thank J. Maldacena for his very useful input on
the following paragraph.}
It falls to $t=0$ for all values of $U$ including $U=0$.
The tension of an anti-D3 brane
placed at $t=0$ is
\eqn\ll{ T_{\overline {D3}}= T_3
H_0^{-1} \left (e^{-\phi_0} +1\right )= {T_3\over \gamma}
{U^2\over e^{-\phi_0} -1}
\ ,}
where the normalization factor is the D3-brane tension
\eqn\threeten{T_3= {1\over 8\pi^3 (\alpha')^2 g_s }\ .}
The plot of this quantity as a function of $U$ is shown in
Figure \antidtri.
For large $|U|$ it again grows as $|U|^{5/4}$. The tension does
not vanish at $U=0$ reflecting the fact that the anti-D3 brane breaks
supersymmetry in the KS background; furthermore, for small $U$
it rises as $\sim U^2$. This means that the scalar mode corresponding
to motion along the baryonic branch has become massive.
Thus, the
non-supersymmetric metastable state of the gauge theory, which is
dual to the anti-D3 brane at the bottom of the KS solution, does
not have a baryonic branch.
For consistency, the massless pseudoscalar Goldstone mode should also be
absent from the spectrum. In fact, it is eaten by the $U(1)$ world volume
gauge field on the anti-D3 brane, which becomes massive.
The term in the world volume gauge theory responsible for this is
\eqn\worldgauge{
\int dA \wedge C_2 =- \int A\wedge F_3\ .}
Since $F_3\sim *(da)$ \GubserQJ, where $a$ is the Goldstone mode,
\worldgauge\ becomes
\eqn\worldgaugenew{
\int A^\mu \partial_\mu a\ ,}
which leads to the Higgs mechanism for the world volume $U(1)$.

\newsec{The D3-brane and a new Approach to Brane Inflation}

The situation is even more interesting for
a D3-brane parallel to
$\IR^{3,1}$. Now the relevant cascading
gauge theory is $SU(1+ M(k+1))\times SU(1+Mk)$. A detailed discussion of
the $k=0$ theory was given in sections 5 and 10, and
of the $k=1$ theory in sections 7 and 10. We will
find that the dual string theory results are in remarkable agreement
with the gauge theory.

The potential of the D3-brane is
\eqn\dthreepot{V(t)= T_3 
H^{-1}(t) (e^{-\phi(t)} -1)
\ .}
The first term comes from
the Born-Infeld term and has a factor of $e^{-\phi(t)}$;
the second term, originating from
the interaction with the background 4-form $C_{0123}$, does not
have this factor. For the KS solution ($U=0$), $\phi(t)=0$ and $V(t)=0$;
therefore, the potential vanishes and the
D3-brane may be located at any point on the deformed conifold.
For
$U\neq 0$ we may use \A\ and \Anought\ to write
\eqn\dthreepotnew{V(t)= {T_3\over
\gamma} {U^2\over  e^{-\phi(t)} +1}
\ .}
Since $\phi(t)$ is a monotonically increasing function,
the D3-brane is attracted to $t=0$.
\fig{Plots of the D3-brane potential
as a function of $t$ for $U=-5$ and $U=-10$.}{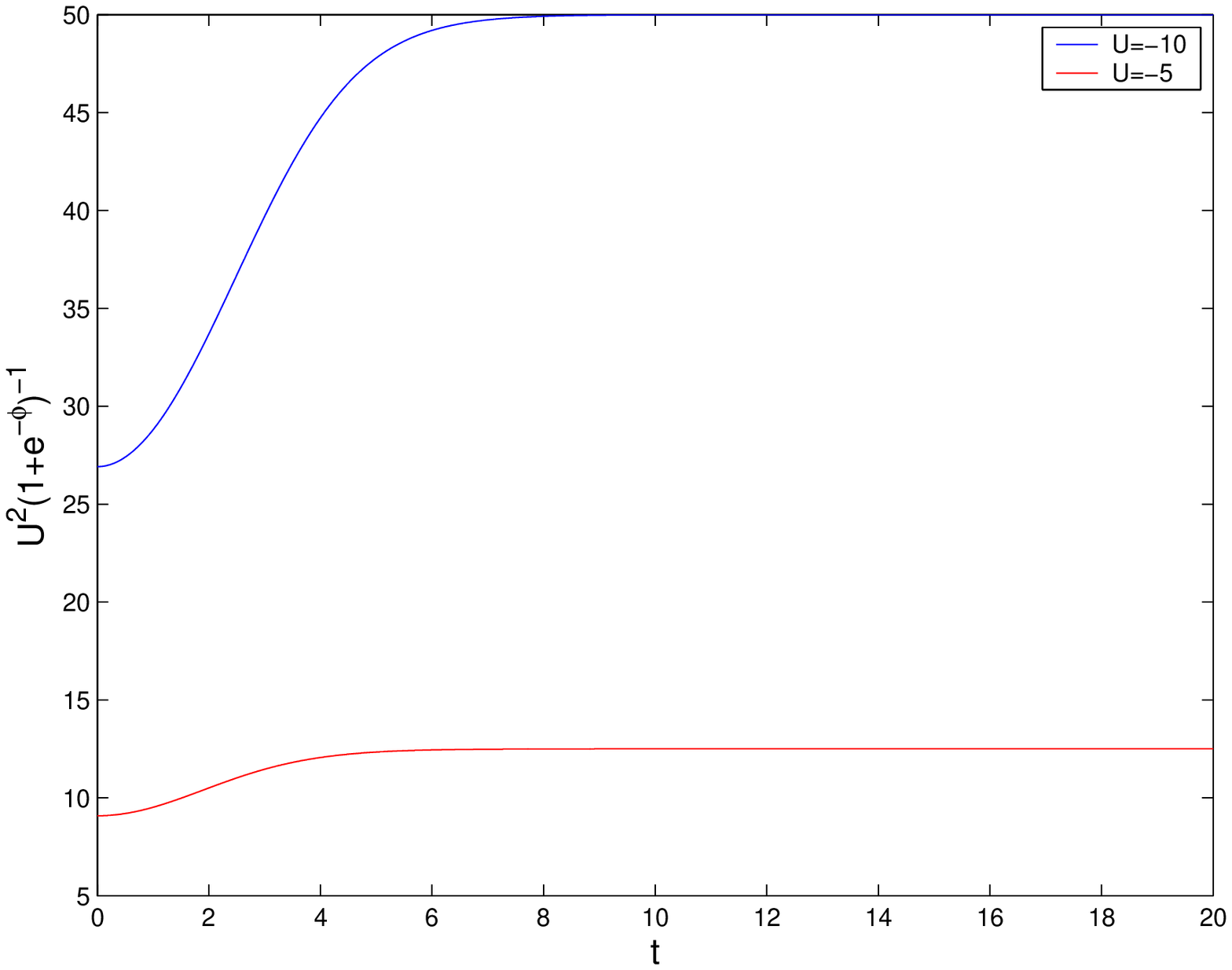}{9cm}\figlabel\prof
Plots of the potential \dthreepot\ for $U=-5$ and $U=-10$ are shown in
Figure \prof. Note that even at $t=0$ the D3-brane has a finite tension
and breaks the supersymmetry.

The fact that the D3-brane has a non-vanishing potential for
a background with $U\neq 0$ follows
from the explicit form of the $10$-dimensional Killing spinor \Butti
\foot{
We thank I. Bena for pointing this out to us.}
\eqn\spinor{\eqalign{
 \Psi &=\alpha \psi+\beta \psi^*\ ,\cr
\alpha &={e^{\phi/8}(1+e^{\phi})^{3/8}\over (1-e^{\phi})^{1/8}}\ ,
\qquad \beta=i{e^{\phi/8}(1-e^{\phi})^{3/8}\over
(1+e^{\phi})^{1/8}} \ .}} The spinor $\psi$ has a definite
4-dimensional chirality, and its charge conjugate $\psi^*$ has the
opposite chirality. At the KS point $\beta=0$, and $\Psi$ has a
definite 4-dimensional chirality. In this case a D3-brane is a
BPS state. But for $U\neq 0$ $\Psi$ does not have a definite
four-dimensional chirality, so none of the supersymmetries of the
background are preserved by the D3-brane.

For small $U$ we may expand
\eqn\dthreepotexp{V(t)= {T_3\over 
8  \gamma}
\left (4 U^2 - 2^{-8/3} I(t) U^4  + O(U^6) \right )
\ .}
Hence,
the attractive force on the D3-brane appears only at order $U^4$.
Note that in the DBI action
the kinetic term for the radial variable $t$ does not have a 
canonical form $\sim \dot t^2$, where $\dot t =\partial t/\partial x^0$. 
Instead, we find the action
\eqn\unusual{
T_3 \left ( f^2 (t) \dot t^2 + H^{-1}(t)  
(e^{-\phi(t)} -1) \right )\ ,
}
where 
\eqn\fsq{ f^2 = {e^{-\phi+x}\over H^{1/2} v}\ .} 
The radial variable $q$ that has the canonical kinetic term may be
found by solving the equation $dq/dt = f(t)\sqrt{T_3}$.
In the asymptotic KT region, $q$ coincides with $\sqrt{T_3} r$
where the standard
variable $r\sim \varepsilon^{2/3} e^{t/3}$.

In models of inflation, one typically defines the parameters\foot{
This is a standard notation in the cosmology literature; this
$\epsilon$ should not be confused with the deformation parameter
of the conifold.} 
\eqn\inflpar{ \epsilon= {M_{Pl}^2\over 2}
V^{-2} \left ({\partial V\over \partial q}\right )^2\ ,\qquad \eta= 
M_{Pl}^2 V^{-1} {\partial^2 V\over \partial q^2}\ , } 
and requires them to be small.
For the potential
\dthreepotexp\ at small $U$ we find that $\epsilon\sim U^4$ and
$\eta\sim U^2$ for all $t$. At large $t$ there is further
suppression of these parameters from the fact that $I(t)$ is
exponentially small: 
this is obvious from the graphs in Figure
\prof. 

For any $U$ we can make $t$ large enough that
$1\gg |\phi(t)|$. Using \phiuv\ we see that this is the case for
\eqn\smallness{ t\gg {3\over 2}\ln |U|\ ;\qquad r^2 \varepsilon^{-4/3}
\gg |U|\ .} 
Then
\eqn\dthreepotnew{
{\gamma\over T_3}V(t)= {U^2\over 1+ e^{-\phi(t)}} \approx
{U^2\over 2} + {U^2\over 4}\phi(t) \approx {U^2\over 2} -
3{U^4\over 256}(4t-1) e^{-4t/3} \ .} Using this expression, we
find \foot{ In the supergravity solution the asymptotic flatness
of the potential is due to the fact that all baryonic branch
backgrounds asymptote to the KT solution \KT, where a D3-brane
experiences no force. This mechanism for generating asymptotically
flat potentials should apply to warped cones more general than the
conifold. For example, for the warped cones over $Y^{p,q}$ found
in \HEK\ the D3-brane is BPS. Resolution of the naked singularity
present at small radius may again lead to variation of the dilaton
and generation of a potential for a D3-brane. But at large radius
the solution has to asymptote to that of \HEK, so the force on a
D3-brane will vanish asymptotically.} 
\eqn\asympflat{-V^{-1} {\partial^2 V\over \partial r^2}\sim
U^2 \varepsilon^{-4/3} (5t-8)e^{-2t}\ . } 
Clearly, for any $U$ this becomes very small at large $t$.
To estimate the range of $U$ for which the slow
roll conditions are obeyed, we need to model a typical compactification.
For this purpose, we introduce a cut-off at a large value
of the radius, $t_{UV}$, where we find the scale of order $(\alpha')^{-1/2}$. 
Since ${\varepsilon^{2/3}\over \alpha'}$ 
is the scale at the bottom of the inflationary throat,
we have\foot{
We will keep track of the exponential terms only, 
and ignore powers of $t$.}
\eqn\scalrelat{
(\alpha')^{-1/2} \sim {\varepsilon^{2/3}\over \alpha'} e^{t_{UV}/3}
\ .
}
It is necessary that $\exp(t_{UV}/3)$ is a large factor: say, $4\times
10^3$ as found in Appendix C of \KKLMMT.
Let us assume that $M_{Pl}^2$ is comparable to the string 
scale $(\alpha')^{-1}$, up to a factor that is not very large
(this is what happens for the numbers adopted in Appendix C of
\KKLMMT).
Then we have 
\eqn\asympflat{|\eta|=- {M_{Pl}^2 \over T_3}
V^{-1} {\partial^2 V\over \partial r^2}\sim
U^2 e^{2t_{UV}/3} e^{-2t}\ . } 
Requiring that this is much smaller than 1 for $t$ around $t_{UV}$ implies
that 
\eqn\reqsmall{ e^{4 t_{UV}/3} \gg U^2 \ .}
This is the same as the requirement that $\phi(t_{UV})$ is close to
zero, \smallness; hence our treatment appears to be self-consistent.
The slow roll condition $1\gg |\eta|$, translated into \reqsmall, leaves
a very large range of $U$ available to modeling of inflation.
 The same is
true for $1\gg \epsilon$. 

Hence, for a D3-brane moving on a resolved warped deformed
conifold
there is no difficulty in achieving very small values of
$\epsilon$ and $|\eta|$ required for the slow-roll inflation. 
This suggests a D-brane inflation model similar to that of KKLMMT
\KKLMMT\ (for earlier ideas in this direction, see \Dvali), but
without the necessity of an anti-D3-brane at $t=0$.
However, one also has to make sure that there are no additional, steeper
corrections to the potential, that are introduced by the compactification
effects.

Let us compare the scale of inflation in our construction
with that in the KKLMMT model. In the KKLMMT model the asymptotic value
of the potential is given by the tension of an anti-D3-brane
placed at $t=0$ in the KS solution, i.e.
\eqn\inflscale{ {2 T_3\over H_{KS}(0)}= 2T_3 {2^{8/3}\over
\gamma I(0)}\approx 8T_3 {2^{2/3}\over 0.71805 \gamma }
\ .}
In our construction, the asymptotic value of the potential
is $T_3 U^2/(2\gamma)$, which is of the same order
as \inflscale\ for $U$ of order $1$.
Note that the large suppression relative to $T_3$ is due to the
factor $\gamma^{-1} \sim e^{-4 t_{uv}/3}$.

For $U\neq 0$ the D3-brane is eventually attracted to $t=0$, and its
tension there is \eqn\dthreetension{ V(0)=  
{T_3\over \gamma} {U^2\over
e^{-\phi_0} +1} \ .} 
The plot of this quantity as a function of
$U$ is shown in Figure \dtri. It vanishes at $U=0$, but for large
$|U|$ it grows as $|U|^{5/4}$. Obviously, the addition of a D3
brane makes the scalar mode massive and lifts the baryonic branch,
in agreement with the conclusions from section 7. The minimum of
the potential is at the $\IZ_2$ symmetric KS point, $U=0$. Even
though $U$ is not a flat direction, let us imagine making $U\neq
0$ by hand, and adding a D3-brane at large $t$. The parameter $U$
induces the FI term in the $U(1)$ gauge theory on the
D3-brane, whose gauge coupling is $g_{YM}$. Hence, roughly, the
flat potential 
\eqn\veryflatpot{
{T_3 U^2\over 2\gamma} ={1\over 2^{2/3} \pi^3 (g_s M)^2}
{U^2\over g_s} {\varepsilon^{8/3}\over (\alpha')^4} }
 at infinity originates from the
${1\over 2} g_{YM}^2 \xi^2$ D-term in this $U(1)$ gauge theory.
Indeed, identifying $\xi$ with $\langle {\cal U} \rangle $ from
\caluvev, we find
\eqn\potcompare{
{1\over 2} g_{YM}^2 \xi^2\sim g_s \xi^2\sim 
(g_s M)^2 {U^2\over g_s} {\varepsilon^{8/3}\over (\alpha')^4 }\ .}
Up to a factor involving powers of $g_s M$, this agrees with
the probe D3-brane result \veryflatpot. This extra factor may appear
in extrapolating from small to
large values of $g_s M$ (see section 11.3 for a discussion of
similar effects).

In the above discussion, the deformation to nonzero $U$ was put in by
hand.  In other words, there is a potential for $U$ which pushes it
to the origin where supersymmetry is restored.  However, when the
throat is embedded into a warped compactification, as in
\Giddings, the $U(1)_{baryon}$ becomes gauged, and we may add a
Fayet-Iliopoulos term $\xi$ for it \GubserTF.
(There has been a lot of discussion about Fayet-Iliopoulos terms in string
theory and supergravity.  Two typical examples are
\refs{\Dine,\BinetruyHH}.)
The Fayet-Iliopoulos term should force the
throat background to a non-zero value of 
$U\sim {\xi (\alpha')^2\over M \varepsilon^{4/3}}$.

Therefore, to construct a real model of D-brane inflation we need
to consider a flux compactification with a non-vanishing $\xi$,
whose throat region is described by a resolved warped deformed
conifold, and add a D3-brane. This makes our construction similar
to the D-term inflation of \refs{\Binetruy,\Halyo}. That model
assumes a $U(1)$ gauge theory coupled to chiral superfields $X$,
$\phi_+$, and $\phi_-$ of charges $0$, $1$ and $-1$, respectively.
Then one adds the superpotential $W=\lambda X\phi_+\phi_-$ and
turns on the Fayet-Iliopoulos term $\xi$. For sufficiently large
$|X|$, and with $\phi_\pm=0$, the potential is found to be
\refs{\Binetruy,\Halyo} \eqn\Dpot{ V_{eff}= {1\over 2} g^2\xi^2
\left ( 1+ {g^2\over 16\pi^2} \ln {\lambda^2 |X|^2\over \Lambda^2}
+ O(g^4)\right ) \ .} This model is similar to our construction
for the $SU(1+ M(k+1))\times SU(1+ Mk)$ gauge theory with the
gauged $U(1)_{baryon}$. In the simplest case of $k=0$ we find
$SU(M+1)\times U(1)_{baryon}$ gauge theory already discussed in
sections 5 and 10. In our construction the meson fields ${\cal
M}_{\alpha \dot\alpha}$ are the analogues of the neutral field
$X$, and the charged fields $A_{\alpha a}$, $B_{\dot\alpha}^b$ are
the analogues of $\phi_\pm$.

We can clearly see
the asymptotically
constant term in the potential in our probe brane
calculation, but we do not observe the logarithmic one-loop
correction.
However, in a flux compactification, various additional
corrections to the potential should appear. In fact, as pointed out in \KKLMMT,
the effects of compactification could make the potential
significantly steeper than what the pure throat limit
\dthreepotnew\ indicates. Investigation of the effects of
compactification is beyond the scope of this paper, but we hope to
address them in the future.

To summarize, our proposal for stringy D-term inflation proceeds
as follows. We consider a warped compactification with fluxes,
which has a warped deformed conifold region. Then we turn on the
Fayet-Iliopoulos term for the gauged $U(1)_{baryon}$ symmetry,
which
forces a breaking of the $\IZ_2$ symmetry. As explained in
\refs{\GubserQJ,\GubserTF} the throat limit of such backgrounds is
provided by the resolved warped deformed conifolds, that were
later constructed in \Butti. Then we add a D3-brane that breaks
supersymmetry, and show that its potential as a function of the
radius varies slowly, at least in the gauge theory limit. Near
$t=0$ the potential gets steeper, and the D3-brane accelerates.
After reaching $t=0$ the D3-brane will undergo oscillations and
internal vibrations which could reheat the Universe. After the
D3-brane stabilizes at $t=0$ in a background with non-vanishing
$U$, it makes a positive contribution to the vacuum energy. This
positive contribution could be used to cancel the negative
contribution to cosmological constant that arises through a
non-perturbative mechanism of the type suggested in \KKLT. As a
result, the net cosmological constant can be made small and
positive, although this may require fine-tuning as usual.
Therefore, our approach appears to avoid the necessity of an
anti-D3-brane \refs{\KKLT, \KKLMMT} or a D7-brane \BurgessIC\
that played 
important roles in earlier constructions;
instead, we use a D3-brane on a {\it resolved}
warped deformed conifold. We leave a detailed investigation of
this model for the future.

\newsec{Discussion}

We have found that the cascading $SU(N_1)\times SU(N_2)$ gauge
theory has many branches of the moduli space. It would be
interesting to extend our systematic study of the moduli space to
more complicated cascading theories, for example to theories on
D-branes near the tip of the cone over $Y^{p,q}$. Some results on
the asymptotic structure of the cascade are already available
\refs{\HEK,\BenvenutiWX}, while in the infrared different
possibilities have been suggested: dynamical SUSY breaking or
runaway behavior where the supersymmetry is restored
\refs{\BerensteinXA\FrancoZU-\BertoliniDI}. In fact, it is
possible that some of the branches of the moduli space lead to
dynamical SUSY breaking while others to runaway behavior. Improved
understanding of these issues should facilitate work on finding IR
completions of the cascading solution found in \HEK.

Another interesting direction raised by our work is to embed
the D3-branes on a resolved warped deformed conifold into a string
compactification with a Fayet-Iliopoulos term. Such a model could be
a useful variation on the KKLMMT model.
These inflationary models have natural generalizations to D3-branes
rolling on other cascading geometries (for example, those found in
\HEK) embedded into flux compactifications.

\bigskip
\centerline{\bf Acknowledgments}
\bigskip

We are indebted to J. Maldacena and G. Moore for collaboration
during the early stages of this project. We are also grateful to
I. Bena and D. Baumann
for very useful discussions. A.~D. would like to thank the
Third Simons Workshop in Mathematics and Physics, where a part of
this work was done. Some of I.~R.~K.'s work on this project took
place at the Aspen Center for Physics, which he thanks for
hospitality. The research of A.~D. is supported in part by grant
RFBR 04-02-16538 and by the National Science Foundation Grant
No.~PHY-0243680. The research of I.~R.~K.\ is supported in part by
the National Science Foundation Grant No.~PHY-0243680 and that of
N.~S.\ by DOE grant \#DE-FG02-90ER40542. Any opinions, findings,
and conclusions or recommendations expressed in this material are
those of the authors and do not necessarily reflect the views of
the National Science Foundation.

\appendix{A}{Review of the Papadopoulos-Tseytlin Ansatz}

The PT ansatz describes
a warped product of the 4-dimensional flat space
$\mathbbR^{3,1}$ and a non-compact six-dimensional manifold $M^6$,
which is roughly speaking the deformed conifold with some internal
warp factors \Pap.

The field strengths are \PT\
(the forms $\epsilon_i$ and $e_i$ are defined below)
\eqn\fieldansatz{\eqalign{
  H_3 =&   h_2(t)  \tilde{\epsilon}_3 \we
(\epsilon_1 \we e_1 + \epsilon_2 \we e_2 )
+  dt \we \big[ h'_1(t)  (\epsilon_1 \we \epsilon_2 + e_1 \we e_2)
 \cr
+&\chi'(t)  (-\epsilon_1 \we \epsilon_2 + e_1 \we e_2)
  + h'_2(t)  (\epsilon_1 \we e_2 -  \epsilon_2 \we e_1 )\big]
 \ ,\cr
F_3 =& P  \te_3\we \big[  \epsilon_1 \we \epsilon_2 +  e_1 \we e_2
-  b(t)  (\epsilon_1 \we e_2 - \epsilon_2 \we e_1) \big]
 \cr
+  &dt \we \big[ b'(t) (\epsilon_1 \we e_1 + \epsilon_2 \we e_2)
\big]\ ,
   \cr
 g_s F_5 &= {\cal F}_5  +  *_{10}{\cal F}_5\ , \ \ \ \ \ \ \
\ \  \ \ \
 {\cal F}_5 = {K}(t) e_1 \we e_2 \we \epsilon_1 \we \epsilon_2 \we \epsilon_3
\ . }} The six-dimensional manifold $M_6$ has the topology of
$\mathbbR^1\times SU(2)\times SU(2)/U(1)=
\mathbbR^1\times\mathbbS^2\times \mathbbS^3$ and the variable $t$
parameterizes the $\mathbbR^1$. The forms $\{e_1,e_2\}$ correspond
to $\mathbbS^2$, while the forms
$\{\epsilon_1,\epsilon_2,\epsilon_3\}$ are the left-invariant
forms on  $\ \mathbbS^3$  as we will see below. The space at
constant $t$ approaches $T^{1,1}$ in the UV region $t \rightarrow
\infty$. In fact, the UV asymptotic metric is $AdS_5\times
T^{1,1}$ modulo slowly-varying logarithms \KT\ which are present
due to the logarithmic RG flow in the dual gauge theory \KN.

The description that
makes the $SU(2)\times SU(2)$ symmetry explicit defines
$M^6$ via an algebraic equation
\eqn\conifold{\eqalign{
&det W =-{\varepsilon^2\over 2}\ ,~~~\ \ \ \ \ \ \ \ \ \ \ \ \ \ \ \ \ \ \ \ \ \ \
W =\rho(t) U_1 Z U_2^+\ , \cr
&U_i=\pmatrix{
   a_i & b_i \cr
  -b_i^* & a_i^* \cr}\in SU(2)\ ,
~~~\ \ \ \ \
Z=\pmatrix{
   0 & \alpha \cr
  \beta & 0 \cr}\ ,\cr
&a_i=\cos(\theta_i/2)e^{i(\psi_i+\phi_i)/2}\ ,~~~\ \ \ \ \ \ \ \
b_i=\cos(\theta_i/2)e^{i(\psi_i-\phi_i)/2}\ ,\cr
 &\rho(t)={\varepsilon e^{-t/2}\over \sqrt{2}}\sqrt{1+e^{2t}}\ ,~~~\alpha ={e^t\over \sqrt{1+e^{2t}}}\ ,\ ~~~\beta={1\over \sqrt{1+e^{2t}}} \ .
}}
Then one gauges the
 $U(1)$ symmetry  that acts by
$\psi_i\rightarrow \psi_i+(-1)^i C$ and introduces
the invariant combination $\psi=\psi_1+\psi_2$
($\psi$ could be also understood as $\psi_2$
when $\psi_1=0$).
Now, we introduce the invariant forms $\epsilon_i$ ($\sigma_i$ is the Pauli matrix)
\eqn\epsilonforms{\eqalign{
2\epsilon_i&=Tr(U_2^+dU_2 \sigma_i)\ ,\cr
\epsilon_1 &\equiv\sin\psi\sin\theta_2 d\phi_2+\cos\psi d\theta_2\ , \cr
\epsilon_2 &\equiv   \cos\psi\sin\theta_2 d\phi_2 - \sin\psi d\theta\ ,\cr
\epsilon_3 &\equiv d\psi + \cos\theta_2 d\phi_2\ ,
}}
and
$
2\hat{\epsilon}_i=Tr(U_1^+dU_1 \sigma_i)
$. The combination
\eqn\ettri{
\tilde{\epsilon}_3=\epsilon_3+\hat{\epsilon}_3=
d\psi+\cos(\theta_1)d\phi_1+\cos(\theta_2)d\phi_2
}
is therefore also invariant under $SU(2)\times SU(2)$.

In the original form \PT, the PT ansatz uses $SU(2)_L$ non-invariant forms
\eqn\forms{
 e_1\equiv d\theta_1\ ,\qquad e_2\equiv  - \sin\theta_1 d\phi_1\ ,  }
 rather than the
invariant $\hat{\epsilon}_1, \hat{\epsilon}_2$.
But $e_1,e_2$ appear only in combinations that could be represented
 via $\hat{\epsilon}_1, \hat{\epsilon}_2$.
For this sake we introduce $SU(2)_R$ non-invariant forms
$  \hat{e}_1\equiv d\theta_2\ , ~~~\hat{e}_2\equiv  - \sin\theta_2 d\phi_2  $
and then express $SU(2)_R$ explicitly  invariant LHS via explicitly
$SU(2)_L$ invariant RHS
\eqn\forms{\eqalign{
&e_1^2+e_2^2=\hat{\epsilon}_1^2+\hat{\epsilon}_2^2\ , \cr
&e_1\epsilon_1+e_2\epsilon_2=\hat{e}_1\hat{\epsilon}_1+\hat{e}_2\hat{\epsilon}_2\ ,\cr
&e_1\wedge \epsilon_1 +e_2\wedge \epsilon_2 = \hat{\epsilon}_1
\wedge\hat{e}_1+ \hat{\epsilon}_2 \wedge \hat{e}_2\ ,\cr
&e_1\wedge \epsilon_2 -e_2\wedge \epsilon_1 =-\hat{e}_1\wedge
\hat{\epsilon}_2 +\hat{e}_2\wedge \hat{\epsilon}_1\ , \cr
&e_1\wedge e_2=-\hat{\epsilon}_1\wedge\hat{\epsilon}_2\ .}}


The PT ansatz is $SU(2)\times SU(2)$ invariant but in general
breaks the $\IZ_2$ symmetry that interchanges $e_1, e_2$ with
$\epsilon_1, \epsilon_2$.
This $\IZ_2$ symmetry is restored for the warped
deformed conifold solution of \KlebanovHB\ by virtue of the
identity $e^g + a^2 e^{-g}= e^{-g}$ as seen from \Pap.

\appendix{B}{First-order equations}

The functions $a,g,x,v,A,h_1,h_2,\chi,K$ depend on the radial
variable $t$ only. The crucial result of \Butti\ is that
supersymmetry of the background requires $a(t)$ and $v(t)$ to
satisfy the coupled first-order equations \eqn\coupled{\eqalign{
a'=&- {{\sqrt{-1 - a^2 - 2\,a\,\cosh t}}\, \left( 1 + a\,\cosh t
\right)\over v \, \sinh t}   -
  {a\,\sinh t\,\left( t + a\,\sinh t \right) \over t\,\cosh t - \sinh t},\cr
v' = & \displaystyle {-3\,a\,\sinh t\over{\sqrt{-1 - a^2 -
2\,a\,\cosh t}}} +\cr
     & + v \, \left[- a^2 \cosh^3 t + 2\,a\,t \coth t +
       a \cosh^2 t \left( 2 - 4\,t \coth t \right) +
       \cosh t \left( 1 + 2\,a^2 \right. \right. \cr
     & \left. \left. - \left( 2 + a^2 \right) t \coth t \right)
           +  \, {t\over \sinh t} \right]/ \left[ \left( 1 + a^2 + 2 a \cosh t
       \right) \left( t \cosh t - \sinh t \right) \right]\ .
}}
The solution of these equations determines other unknown
functions through \phieq\ and
\eqn\K{\eqalign{
& e^{2g}=-1-a^2+2aC\ ,\cr
& e^{2x}= \left({g_s M\alpha'\over 2}\right)^2
{(bC-1)^2\over 4(aC-1)^2} e^{2g+2\phi}(1-
e^{2\phi})\ , \cr
& b=-{t\over \sinh(t)}\ ,\cr & h_1=-C h_2\ ,\cr
& h_2= \left({g_s M\alpha'\over 2}\right){e^{2\phi}(bC-1)\over 2S}\ ,\cr
& \chi'=\left({g_s M\alpha'\over 2}\right)
a(b-C)(aC-1)e^{2(\phi-g)}\ ,\cr &
K=- \left({g_s M\alpha'\over 2}\right)
(h_1+bh_2)=
\left({g_s M\alpha'\over 2}\right)^2 e^{2\phi} {(bC-1)(C-b)\over 2S}
\ ,  \cr &P=-\left({M\alpha'\over 4}\right) , }} where
$C=-\cosh(t),\ S=-\sinh(t)$.

We have fixed normalizations from the condition that there
are $M$ units of flux of the RR
$3$-form field strength $F_3$ through the
$\mathbbS^3$
\eqn\ll{ {1\over 4\pi^2
\alpha'}\int_{\mathbbS^3}F_3=M\ . }
The integer $M$ is dual to the
difference between the numbers of colors of the two gauge groups.

Since \coupled\ is a system of two coupled first-order equations,
one might expect a two-parameter family of solutions, but in fact
all solutions regular at $t=0$ are parameterized by just one real
parameter $y$. The small $t$ expansion found in \Butti\ is
\eqn\ave{\eqalign{
& a=-1+\left({1\over 2}+{y\over 3}\right) t^2+...\ ,\cr
& v=t+\left(-{13\over 96}+{17 y^2\over 216}\right)t^3+... }}
The parameter $\xi_{BGMPZ}$ defined as $\xi_{BGMPZ}=1/2+y/3$
varies from $1/6$ to $5/6$ when $y$ varies from $-1$ to $1$
along the baryonic branch. Any value
$\xi_{BGMPZ}$ is related to $1-\xi_{BGMPZ}$ by the $\IZ_2$ symmetry $y\rightarrow -y$
which changes the sign of $U$
but leaves $v(t)$ and the combination $a(t)e^{-g(t)}$
invariant.
The $\IZ_2$ symmetric value $y=0$ obviously corresponds to the
KS solution dual to the locus on the baryonic branch where $|{\cal
A}|= |{\cal B}|= \Lambda_1^{2M}$ and $e^{2g}+a^2=1$.

\listrefs

\end